\documentclass[useAMS,usenatbib]{mn2e}
\usepackage{epsfig}
\title[Properties of dust and detection of H$\alpha$ emission in LDN 1780]{Properties of dust and detection of H$\alpha$ emission in LDN1780}
\author[C. del Burgo]{C. del Burgo$^{1}$\thanks{E-mail:
cburgo@cp.dias.ie (CdB)}, L. Cambr\'esy$^2$
\\
$^{1}$Dunsink Observatory, Dublin Institute for Advanced Studies, Castleknock, 
Dublin 15, Ireland\\
$^{2}$Observatoire Astronomique de Strasbourg, F-67000 Strasbourg, France\\ 
}
\begin{document}
\date{Accepted 2005 XXXX XX. Received 2005 XXXX XX; in original form 2005 XXXX 
XX}

\pagerange{\pageref{firstpage}--\pageref{lastpage}} \pubyear{2002}

\maketitle

\label{firstpage}

\begin{abstract}
  We present {\em ISOPHOT\/} observations between 60 and 200 $\mu$m 
  and a near-infrared extinction map of the small intermediate-density cloud 
  LDN 1780 (Galactic coordinates l=359$^{\rm o}$ and b = 36.8$^{\rm o}$). 
  For an angular resolution 
  of 4$^\prime$, the visual extinction maximum is $A_V=4.4$ mag. 
  We have used the {\em ISOPHOT\/} data together with 
  the 25, 60 and 100 $\mu$m {\em IRIS\/} maps (Miville-Desch\^{e}nes \& Lagache 2005)
  to disentangle the {\em warm} and {\em cold} components of large dust grains that
  are observed in translucent clouds (Cambr\'esy et al. 2001, del Burgo et al. 2003) 
  and dense clouds (del Burgo \& Laureijs 2005).
  The warm and cold components in LDN 1780 have different properties 
  (temperature, emissivity) and spatial distributions, 
  with the warm component surrounding the cold component. The warm component
  is mainly in the illuminated side of the cloud facing the Galactic plane and 
  the Scorpius-Centaurus OB association, as in the case of
  the HI excess emission (Mattila \& Sandell 1979). The cold component is 
  associated with the $^{13}$CO (J=1-0) line integrated $(W_{13})$, which trace molecular gas 
  at densities of $\sim$10$^{3}$ cm$^{-3}$. The warm component has a uniform colour temperature
  of 25$\pm$1\,K (assuming $\beta=2$), and the colour temperature of the
  cold component slightly varies between 15.8 and 17.3\,K ($\beta=2$, ${\Delta}$T=0.5\,K).
  The ratio between the emission at 200 $\mu$m of the cold component
  ($I^c_\nu(200)$) and $A_V$ is $I^c_\nu(200)/A_V$=12.1$\pm$0.7 MJy sr$^{-1}$ mag$^{-1}$
  and the average ratio $\tau_{200}/A_V=(2.0\pm 0.2) \times 10^{-4}$ mag$^{-1}$. The far infrared
  emissivity of the warm component is significantly lower than that of the cold component.
  The H$\alpha$ emission ($I_{\nu}({\rmn H}\alpha)$) and $A_V$ correlate
  very well; a ratio $I_{\nu}({\rmn H}\alpha)/A_V$ = 2.2$\pm$0.1 Rayleigh mag$^{-1}$
  is observed. This correlation is observed for a relatively large range of column 
  densities and indicates the presence of a source of ionisation that can penetrate
  deep into the cloud (reaching zones with optical extinctions $A_V$ of 2 mag).
  Based on modelling predictions we reject out a shock front as precursor
  of the observed H$\alpha$ surface brightness although that process could
  be responsible of the formation of LDN 1780. Using the ratio $I_{\nu}({\rmn H}\alpha)/A_V$
  we have estimated a ionisation rate for LDN 1780 that results to be 
  $\sim$10$^{-16}$ $\gamma$ s$^{-1}$. We interpret this relatively high value as due
  to an enhanced cosmic ray flux of $\sim$10 times the standard value. 
  This is the first time such an enhancement
  is observed in a moderately dense molecular cloud.
  The enhancement in the ionisation rate could be explained
  as result of a confinement of low energy ($\sim$100 MeV) cosmic rays 
  by self generated MHD waves (Padoan \& Scalo 2005). The origin of
  the cosmic rays could be from supernovae in the Scorpio-Centaurus OB 
  association and/or the runaway $\zeta$ Ophiuchus.
  The observed low $^{13}$CO abundance and relatively high temperatures of the 
  dust in LDN 1780 support the existence of a heating source that can
  come in through the denser regions of the cloud. 
\end{abstract}


\section{Introduction}

The Lynds Dark Nebula LDN 1778/1780 (Lynds 1962) is a moderately dense 
(translucent) region with Galactic coordinates l = 359$^{\rm o}$ and b=36$^{\rm o}$.8. This 
compact nebula, hereinafter only referred as LDN 1780, is located at
h $\approx$ h$_\odot$ (=15 pc) + 66 pc (assuming a distance of $110\pm10$ pc,
Franco 1989) from the Galactic midplane. It is 1.28 pc $\times$ 0.96 pc in size, slightly
elongated in the E-W direction on the sky. It shows a well defined boundary
towards NW and is more diffuse on the SE side, which faces the Galactic Plane.

LDN 1780 was studied together with other nearby clouds (LDN 134, LDN 183, LDN 169, MBM 38 and 
MBM 39) by Laureijs et al. (1995, hereinafter LFHMIC95) from carbon monoxide
observations at 2.7 mm of CO(J=1-0), $^{13}$CO(J=1-0) and
$^{18}$CO(J=1-0)\footnote{No detection of $^{18}$CO(J=1-0) in LDN 1780
was achieved.}, 12, 25, 60 and 100 $\mu$m {\em IRAS} data, and blue
extinction ($A_B$) from star counts. They found that the $^{13}$CO(J=1-0) 
abundance in LDN 1780 is $\approx$4 times lower than in the opaque clouds
LDN 134, LDN 183 and LDN 169. The 100 $\mu$m emission excess
${\Delta}I_\nu(100) \equiv I_\nu(100)-I_\nu(60)/{\Theta}$ (where $I_\nu(60)$
and $I_\nu(100)$ are respectively the emissions at 60 and 100 $\mu$m and
$\Theta$ is the average ratio $I_\nu(100)/I_\nu(60)$, 
which is 0.27 for LDN 1780 and 0.21 for the rest of the clouds) corresponds to 
the emission from the ``classical'' big grains (without the contribution from
the very small grains). LFHMIC95 found that there is a
good correlation of ${\Delta}I_\nu(100)$ and the $^{13}$CO(J=1-0) line integrated 
($W_{13}$). Although the ratio ${\Delta}I_\nu(100)/W_{13}$ in LDN 1780 
is significantly higher than in the other clouds, it presents a 
similar ratio ${\Delta}I_\nu(100)/A_B$. LFHMIC95 concluded that the anisotropic 
UV radiation field in the complex affects the density distribution, especially 
on the illuminated side of the clouds with a bigger halo, but also in the
densest, well-shielded regions, shifted with respect to the cloud centres to
the opposite side of the illuminated halos. Only LDN 1780 lacks a diffuse halo
and the radiation field is still affecting the molecular content given the low
opacity of the cloud (LFHMIC95).

LDN 1780 is near Radio Loop I (Berhuijsen 1971), likely just at the boundary
of the Local Bubble (a Galactic region of radii 65-150 pc around the Sun of
low density and hot gas with a temperature of 10$^6$ K; see Lallement et al. 2003 and references
therein). Radio Loop I is thought to be the limb of the neighbouring hot
bubble Loop I, a superbubble centred on the Sco-Cen OB association 
(distance of 120-160 pc, de Geus, de Zeeuw \& Lub 1989) that is expanding at 
10-20 km s$^{-1}$ due to stellar winds and supernovae (de Geus 1992).
X-ray observations at $\frac{1}{4}$, $\frac{3}{4}$ and 1.5 KeV
support that LDN 1780 is located at the boundary of the Local Bubble and show 
that the cloud is not affected by any possible hot gas flowing from the hotter 
bubble Loop I to the Local Bubble (Kuntz, Snowden \& Verter, 1997).

LDN 1780 is the only translucent cloud in which has been observed 
Extended Red Emission (ERE) (Chlewicki \& Laureijs 1987) so far. The
ERE in this cloud has a broad, unstructured band, with a peak near 7000 \AA.
This peak's wavelength is longer than expected,
and it could be due to silicon nanoparticles in a environment with relatively high
electron density and lower UV photon density in comparison with
the diffuse interstellar medium (Smith \& Witt 2002). 

Observations in the 21 cm HI line and the 18 cm OH lines of LDN 1780 were 
carried out using the Effelsberg 100 m telescope by Mattila \& Sandell (1978). 
They concluded that the atomic and molecular gas are spatially connected with 
each other and the dust cloud. Observations of LDN\ 1780 at 9 cm CH (Mattila
1986), 6 cm H$_2$CO (Clark \& Johnson 1981) have been performed too.
Molecular and HI excess emission spectra show a peak intensity at the 
local-standard-of-rest velocity v$_{LSR}$ = 3.5 km s$^{-1}$ and broad blue 
shifted profiles. Based on CO(J=1-0), $^{13}$CO(J=1-0) and C$^{18}$O(J=1-0) 
observations T\'oth et al. (1995)\footnote{A faint C$^{18}$O(J=1-0) emission 
spectra were obtained in few positions only.} (hereinafter THLM95)
claimed the observed cometary like structure and morphology of LDN 1780 is 
result of shock fronts from the Sco-Cen OB association, as well as an
asymmetric UV radiation from the Galactic disk and the Upper Scorpius group.

LDN 1780 presents\footnote{A distance of 110 pc for LDN 1780 was assumed to 
determine the following parameters.} a unique $^{13}$CO(J=1-0) core with a 
regular morphology (LFHMIC95, THLM95), a narrow velocity component of 
$\Delta{V}\approx$0.6 km s$^{-1}$ (LFHMIC95; 0.52 km s$^{-1}$ from THLM95), an 
average density of 10$^{3}$ cm$^{-3}$ (LFHMIC95; a lower value of 0.6 10$^{3}$ 
cm$^{-3}$ is obtained from its column density and size determined by THLM95), 
and a total mass of almost 10 $M_{\odot}$ (LFHMIC95; 8.3 $M_{\odot}$ according 
to THLM95). It displays a $r^{-2}$ density distribution (LFHMIC95) and it is in 
virial equilibrium (THLM95). The total mass in the whole $^{13}$CO(J=1-0) 
emitting region is 18 $M_{\odot}$ (THLM95), and the estimated mass of the cloud 
envelope is 1.8-3.6 $M_{\odot}$ (THLM95). The total mass of LDN 1780 is then 
$\approx$20 $M_{\odot}$.

In this paper we present observations of LDN 1780 obtained using
{\em ISOPHOT\/} in the spectral range 60-200 $\mu$m that trace the emission
from very small grains (VSGs) and big grains (BGs). These far-infrared data have been
complemented with Two Micron All Sky Survey (2MASS) near-infrared photometry
(Cutri et al. 2003), the 25, 60 and 100 $\mu$m {\em IRIS\/} maps
(Miville-Desch\^{e}nes \& Lagache 2005), H$\alpha$ maps (Finkbeiner 2003), 
and CO molecular line measurements (LFHMIC95). We have studied the distribution and 
the properties (temperature and emissivity) of the dust grains in the cloud, 
and compared some of these properties with the molecular gas emission, the
H$\alpha$ emission and their distributions. An extinction map is derived from 
2MASS photometry, and converted to optical extinction $A_V$ assuming a standard 
extinction law. This map is used as a tracer of the column density of dust at 
short wavelengths. The warm and cold components of large grains (see del Burgo
et al. 2003, hereinafter Paper I) are separated following a similar 
method to that used by del Burgo \& Laureijs (2005) (hereinafter Paper II).
The extinction map is compared with the emission and the optical depth at 200 
$\mu$m of the cold component, to investigate emissivity variations of the big
dust grains across the cloud, as carried out in previous studies (PaperI; Paper II). 
The H$\alpha$ surface brightness in the cloud is corrected from 
extinction, and a cosmic ionisation rate is derived from the ratio 
$I_{\nu}({\rmn H}{\alpha})/A_V$.
We discuss the nature of the ionisation source to explain the presence of 
H$\alpha$ emission in moderately extinguished regions ($A_V\sim$2 mag) in the cloud.
We have also studied the correlations of the far infrared dust tracers with 
molecular gas measurements of CO at 2.7 mm. The results for this intermediate density
cloud are compared with other studies of translucent and dense regions.
Section 2 describes the data processing. Sections 3 and 4 are respectively
devoted to the properties of the two big dust grain components and the ionisation
process in the cloud. Finally, Section 5 presents the conclusions.

\section{Acquisition, processing and preparation of the data}

\subsection{{\em ISOPHOT\/} observations}

A nearly squared $38^{\prime}\times 38^{\prime}$ (equivalent to 
1.2 pc $\times $1.2 pc for a distance of 110 pc) region centred at the
equatorial coordinates $\alpha_{2000}$=15$^h$40$^m$09.5$^s$ and 
$\delta_{2000}$=-7$^{\rm o}$12$^{\prime}$28.7$^{\prime\prime}$ and a
stripe-shaped region centred at $\alpha_{2000}\approx$15$^h$39$^m$50$^s$ 
and $\delta_{2000}\approx$-7$^{\rm o}$11$^{\prime}$, were mapped with
{\em ISOPHOT\/} (Lemke et al. 1996), an instrument on board of ESA's
Infrared Space Observatory ({\em ISO}, Kessler et al. 1996). The stripe-like region
($\sim$38 arcmin along E-W; $\sim$3 arcmin along N-S) was observed using
the Astronomical Observation Template PHT22 in raster mapping mode with
the array detectors C100 (3$\times$3 pixels; 46$''$ pixel$^{-1}$) and the
60, 70 and 100 $\mu$m filterbands, and also with C200 (2$\times$2 pixels;
92$''$ pixel$^{-1}$) and the 120, 150 and 200 $\mu$m filterbands. The squared
region was observed in the same observing mode with C100 at 100 $\mu$m and
 with C200 at 200 $\mu$m, and it nearly covers LDN 1780. For an overview of the
different observing modes see the {\em ISOPHOT\/} Handbook (Laureijs et al.
2003).

Table \ref{tab:sample} lists the log of the observations, with the {\em ISO} 
identification ({\em ISO}$_{id}$), the equatorial coordinates (right ascension
and declination J2000.0) of the raster centre, the reference wavelength of the
{\em ISOPHOT} filter band ($\lambda_{ref}$), the raster dimensions
(N$\times$M), the distances between adjacent points and lines (dN and dM,
respectively), and the exposure time (t$_{exp}$) per sky position. The
redundancy is generally 2 or 1. There are also some small gaps that
are regularly interleaved along the stripe-like region when surveyed with 
the C100 array detector. The total exposure time per 
pixel is obtained multiplying the redundancy and t$_{exp}$. In addition to 
the observations of the object, two measurements of the internal Fine 
Calibration Source (FCS) were obtained with the same filter, detector and 
aperture just before and after each map.

\begin{table*}\begin{center}\begin{footnotesize}
\caption{{\em ISOPHOT\/} observations of LDN 1780. \label{tab:sample}}
\begin{tabular}{lllcccccc}
\hline
$\alpha_{\rm J2000}$ & $\delta_{\rm J2000}$ & $ISO_{\rm id}$ & $\lambda_{\rm ref}$ & N$\times$M & dN, dM & Area & t$_{exp}$ & Remark \\
 $[hh~mm~ss]$ & $[^{\rm o~~\prime~~\prime\prime}]$ &  & [$\mu$m]  &  & [arcsec, arcsec] & [arcmin$^2$] & [s] & \\ \hline
 15 39 40.93 & -7 10 35.04 & 43100206 &  60 & 16$\times$2  & 180,  90 &  206 & 34 & stripe \\
 15 39 40.93 & -7 10 35.02 & 43100207 &  70 & 16$\times$2  & 180,  90 &  206 & 34 & stripe \\
 15 39 40.93 & -7 10 35.01 & 43100208 & 100 & 16$\times$2  & 180,  90 &  206 & 34 & stripe \\
 15 39 58.55 & -7 11 39.14 & 43100209 & 120 & 13$\times$2  & 180,  90 &  249 & 34 & stripe \\
 15 39 58.55 & -7 11 39.12 & 43100212 & 150 & 13$\times$2  & 180,  90 &  249 & 34 & stripe \\
 15 39 58.55 & -7 11 39.11 & 43100210 & 200 & 13$\times$2  & 180,  90 &  249 & 34 & stripe \\
 15 40 09.50 & -7 12 28.74 & 43100630 & 100 & 18$\times$24 & 135,  90 & 1495 & 16 & square \\
 15 40 09.51 & -7 12 28.69 & 43100629 & 200 & 13$\times$13 & 180, 180 & 1469 & 17 & square \\
\hline
\end{tabular}\end{footnotesize}\end{center}
\end{table*}

The data reduction was performed with the {\em ISOPHOT\/} Interactive Analysis 
software {\small PIA} V10.0 (Gabriel et al. 1997). Signals corresponding to
the same sky position were averaged after correcting for glitches. The slow
response or transient behaviour of detector C200 for flux changes is generally visible
only on the first few raster steps of the maps, but it did not affect the main
features of these maps. Memory effects due to crossing bright sources are
negligible. All standard signal correction steps (reset interval correction, 
dark signal subtraction, signal linearisation) were applied. For flux 
calibration the detector's actual response was determined from the
second FCS measurement, which is observed just after the object. Unless
explicitly stated no colour correction was applied on the data.

We do not correct from the by-passing sky light (del Burgo, Heraudeau,
\'Abr\'aham 2003). Note, however, our analysis minimises the impact of any
undesirable systematic offset, for instance, using pixel-pixel correlations
to derive average emission ratios (see Paper I).

The first quartile normalisation flat-fielding method of {\small PIA} was used 
to correct for the remaining responsivity differences of the individual
detector pixels.

\begin{figure*}
\vspace*{-1.5cm}\centerline{
\hspace*{0.1cm}\psfig{figure=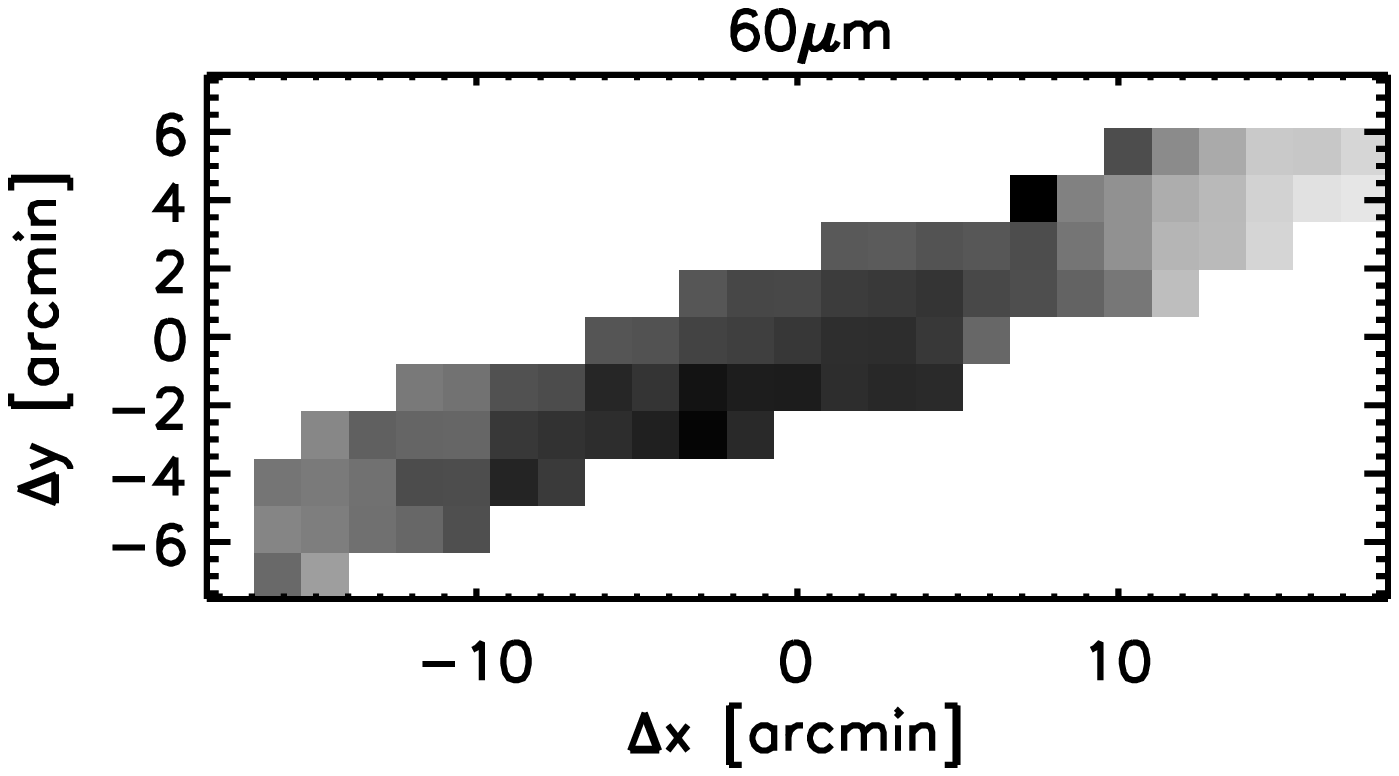,width=85mm,angle=0}
\hspace*{0.1cm}\psfig{figure=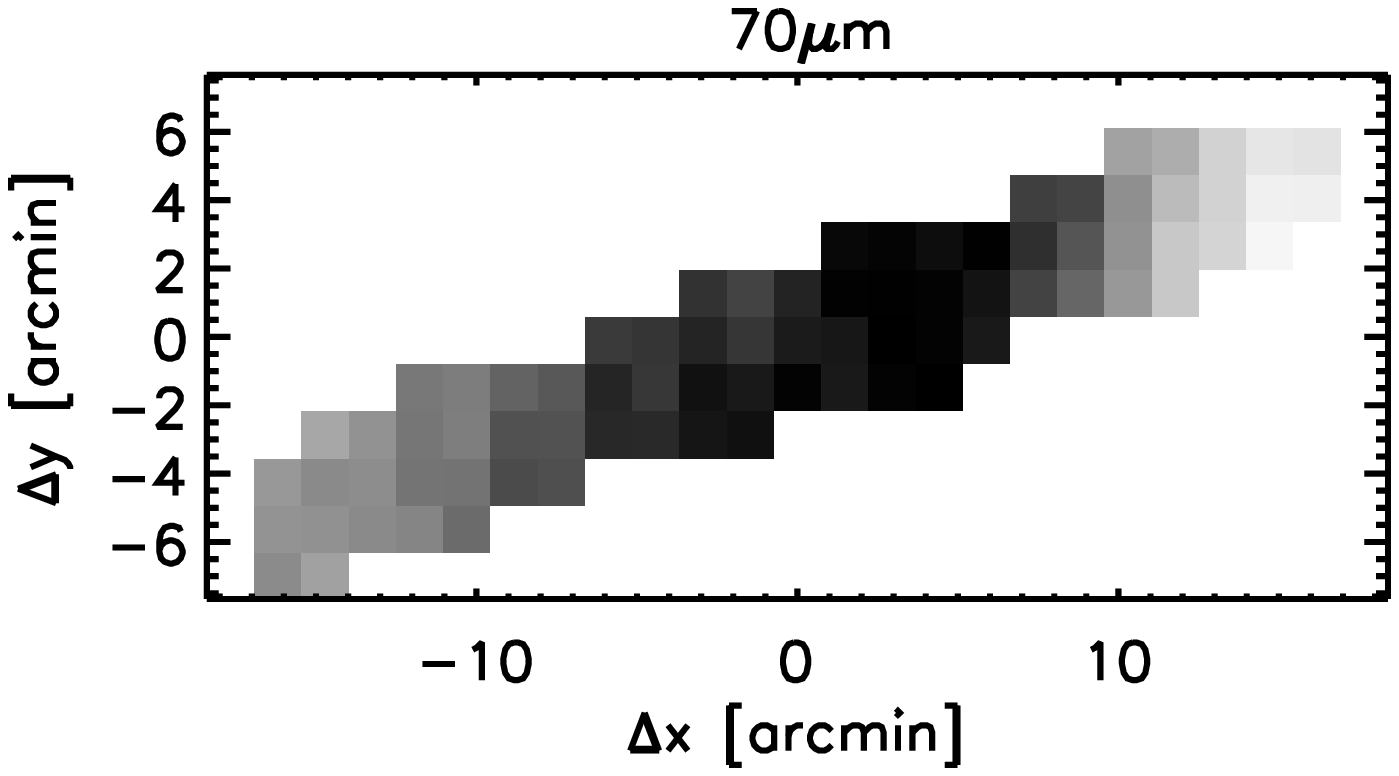,width=85mm,angle=0}}
\vspace*{-3.75cm}\centerline{
\hspace*{0.1cm}\psfig{figure=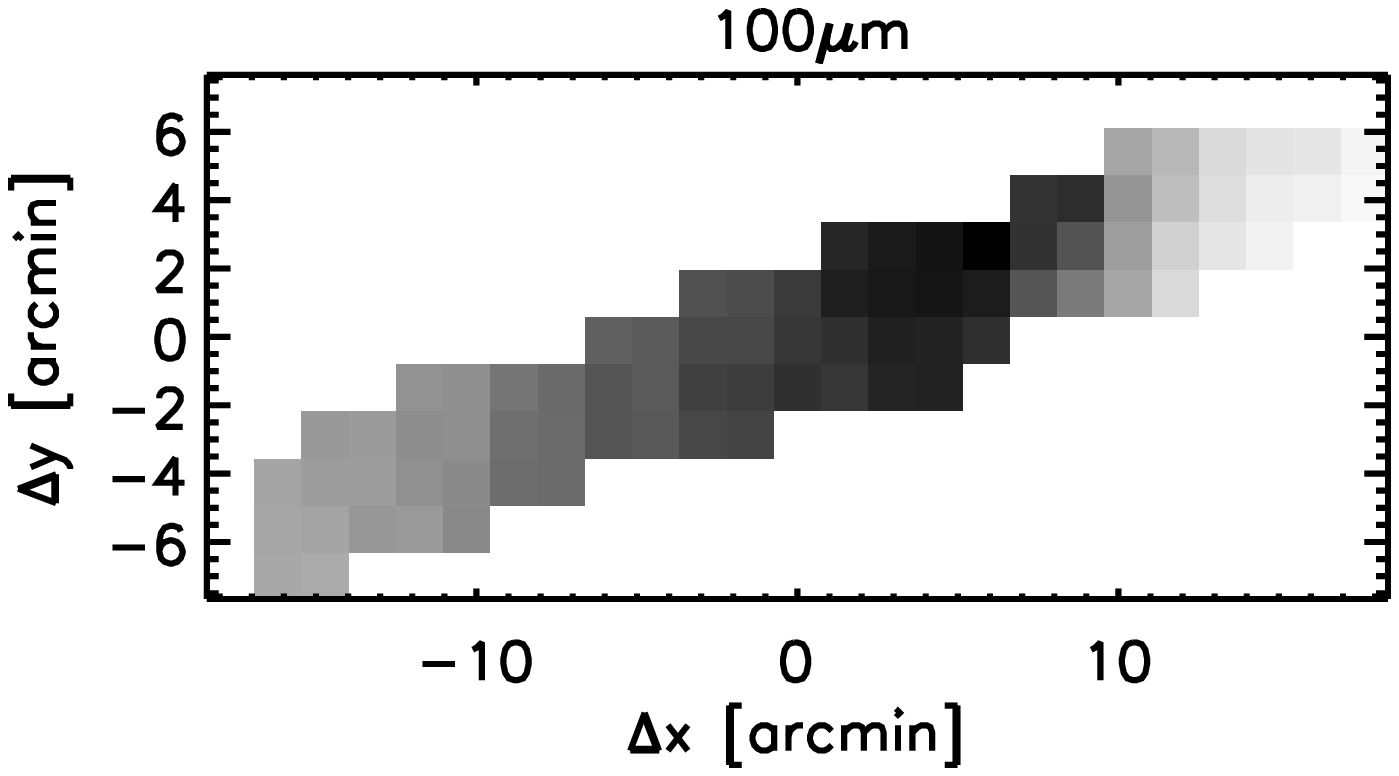,width=85mm,angle=0}
\hspace*{0.1cm}\psfig{figure=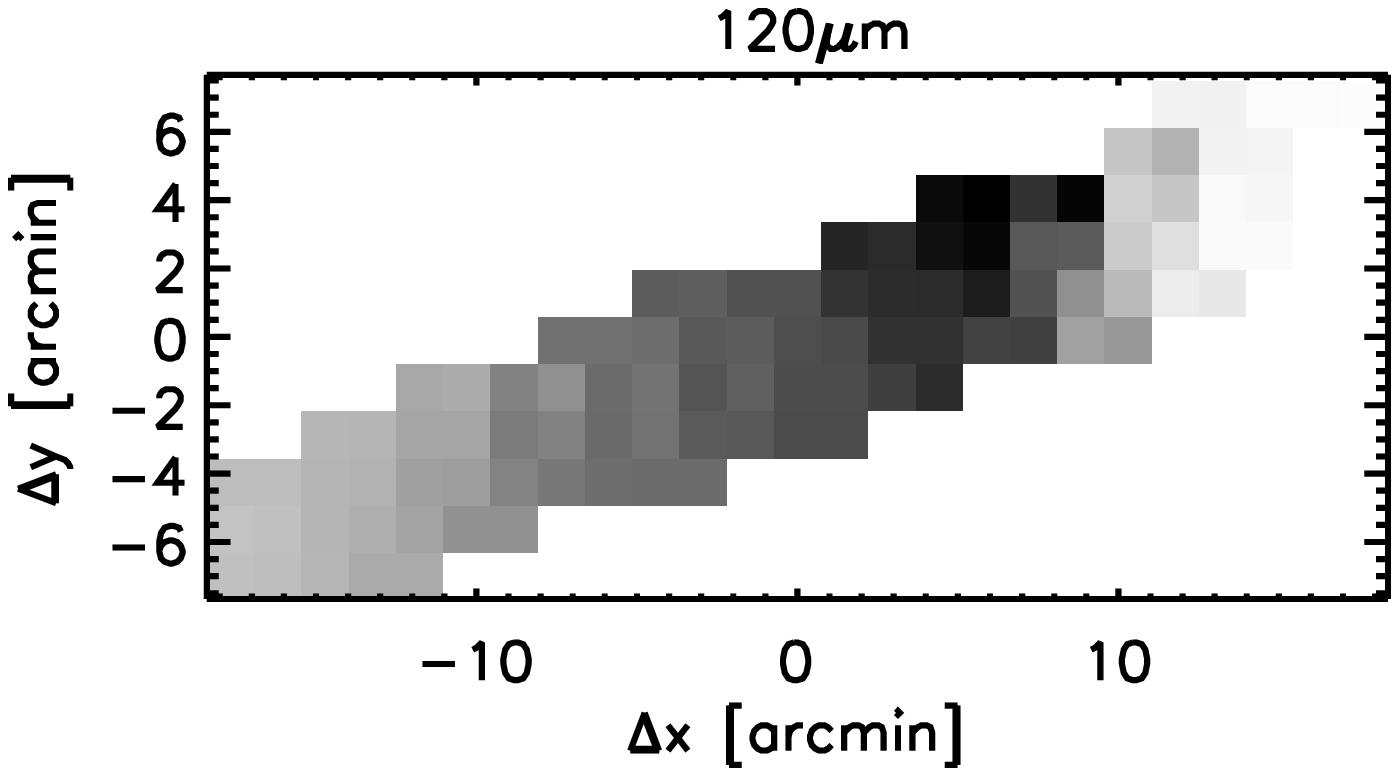,width=85mm,angle=0}}
\vspace*{-3.75cm}\centerline{
\hspace*{0.1cm}\psfig{figure=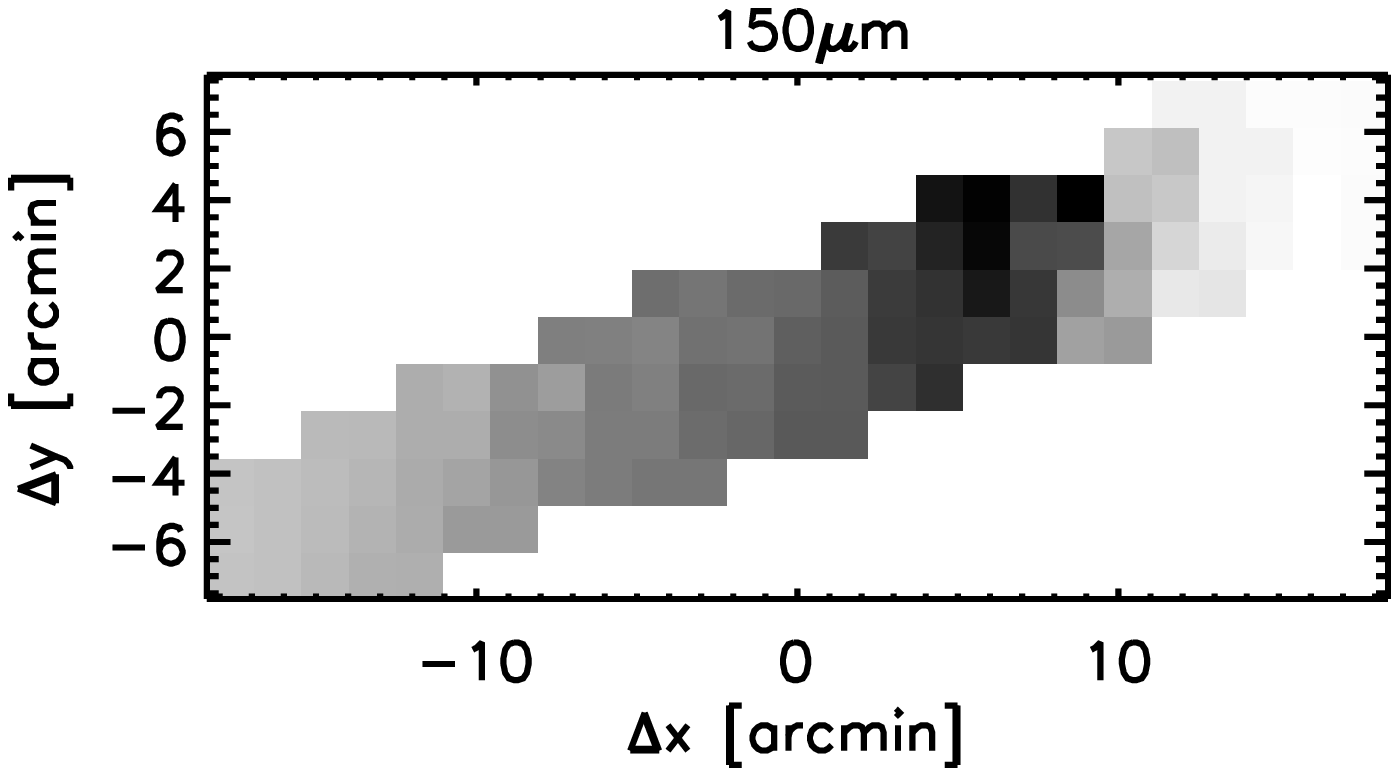,width=85mm,angle=0}
\hspace*{0.1cm}\psfig{figure=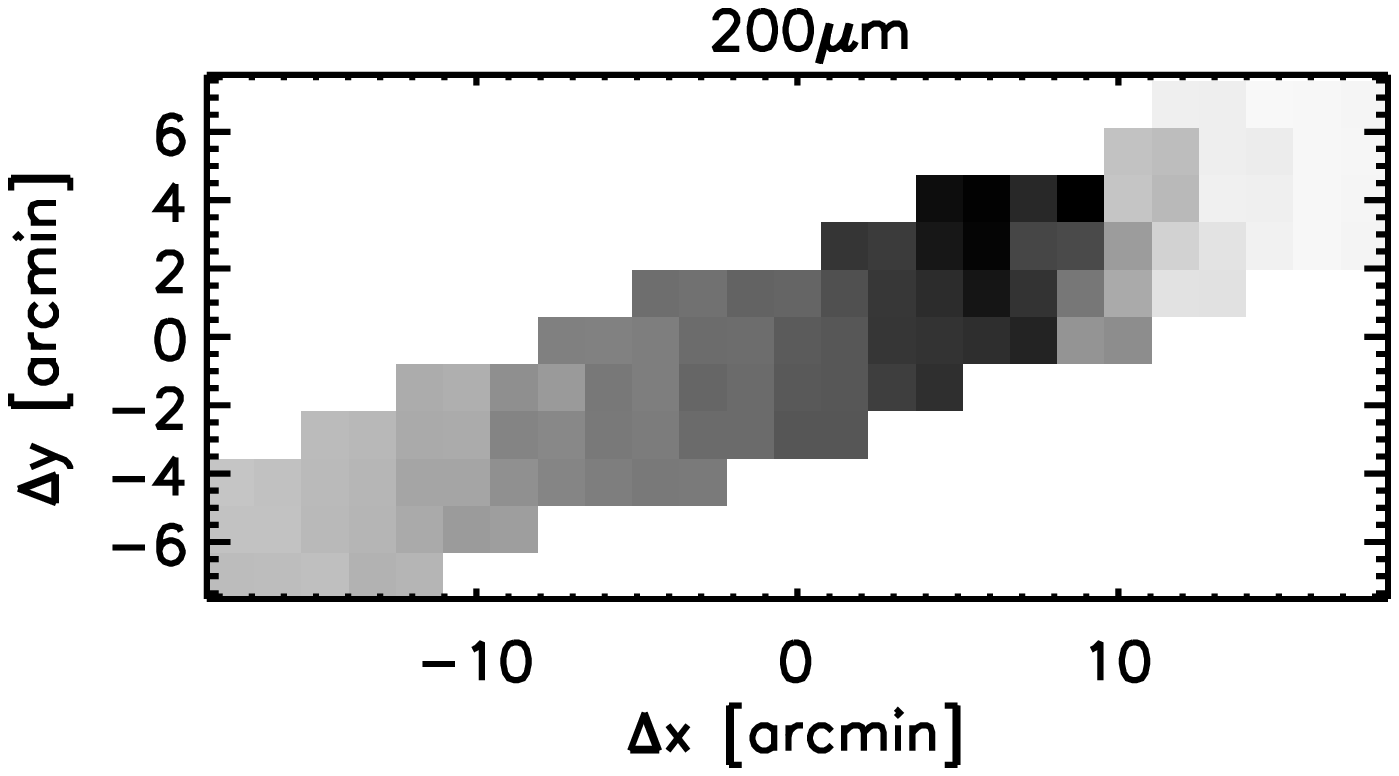,width=85mm,angle=0}}
\vspace*{-3.75cm}\centerline{
\hspace*{0.1cm}\psfig{figure=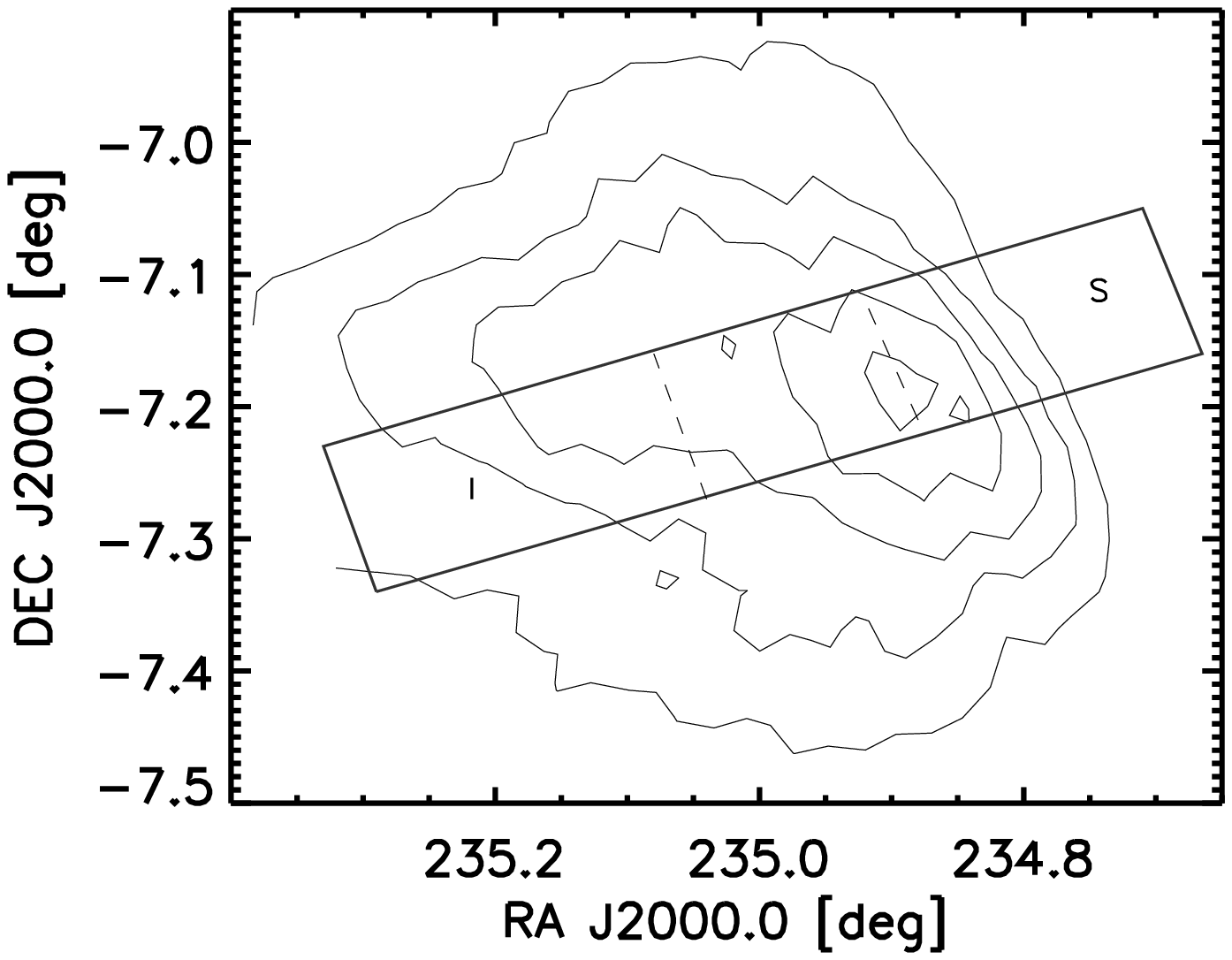,width=85mm,angle=0}}
\vspace*{0.3cm}
\caption{From {\em top} to {\em bottom}, from {\em left} to {\em right}: 
Emission at 60, 70, 100, 120, 150 and 200 $\mu$m smoothed to the angular 
resolution the 200 $\mu$m emission corresponding to the stripe region, which is 
overplotted on the 200 $\mu$m emission map corresponding to the square region. 
{\em I} and {\em S} labels respectively stands for the illuminated and shadowed
regions separated by dashed lines from the rest of the stripe (see Sect. 2.4.1).}
\label{fig:figure1}
\end{figure*}

\begin{figure*}
\vspace*{-0.3cm}\centerline{
\hspace*{0.1cm}\psfig{figure=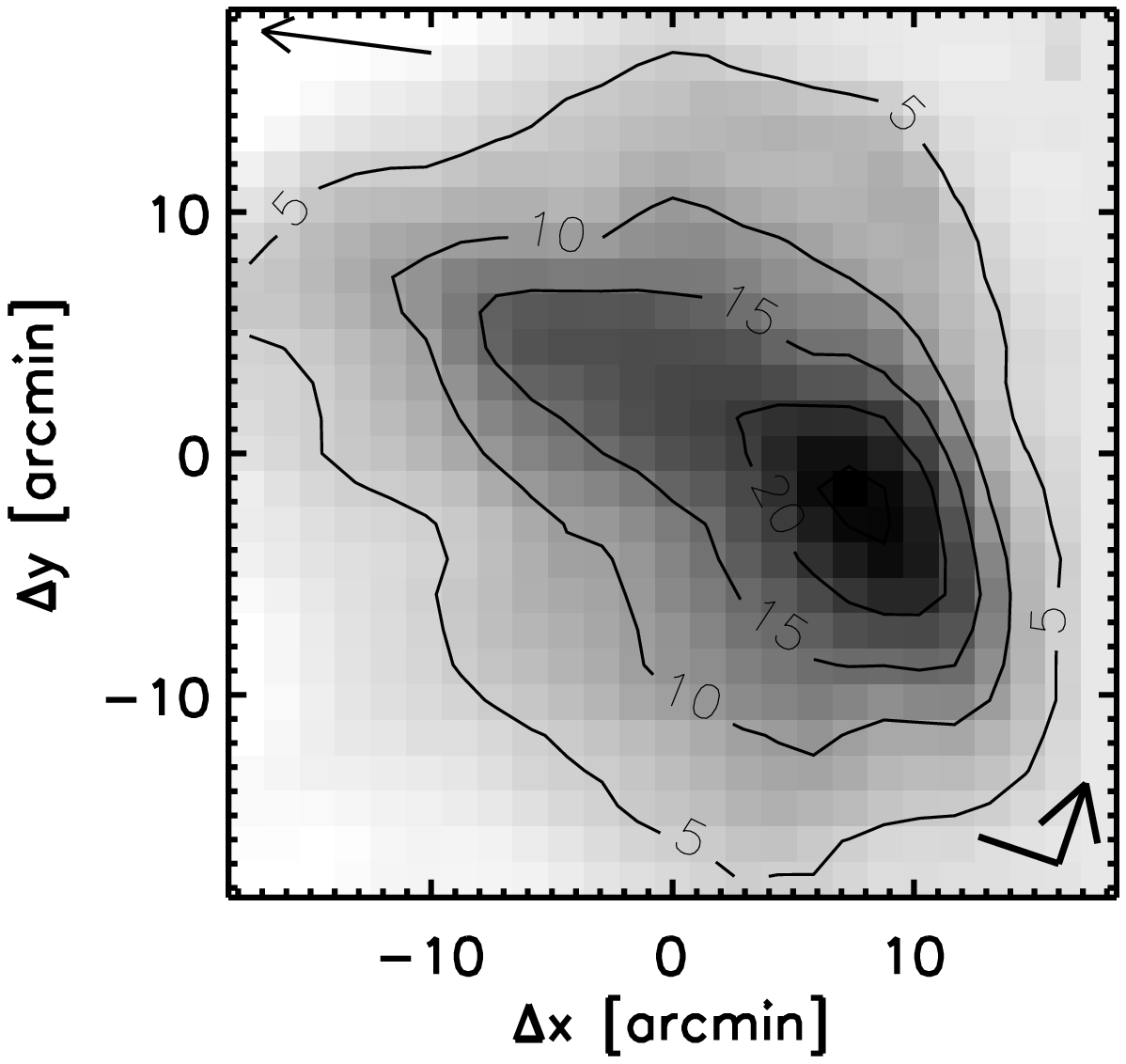,width=85mm,angle=0}
\hspace*{0.1cm}\psfig{figure=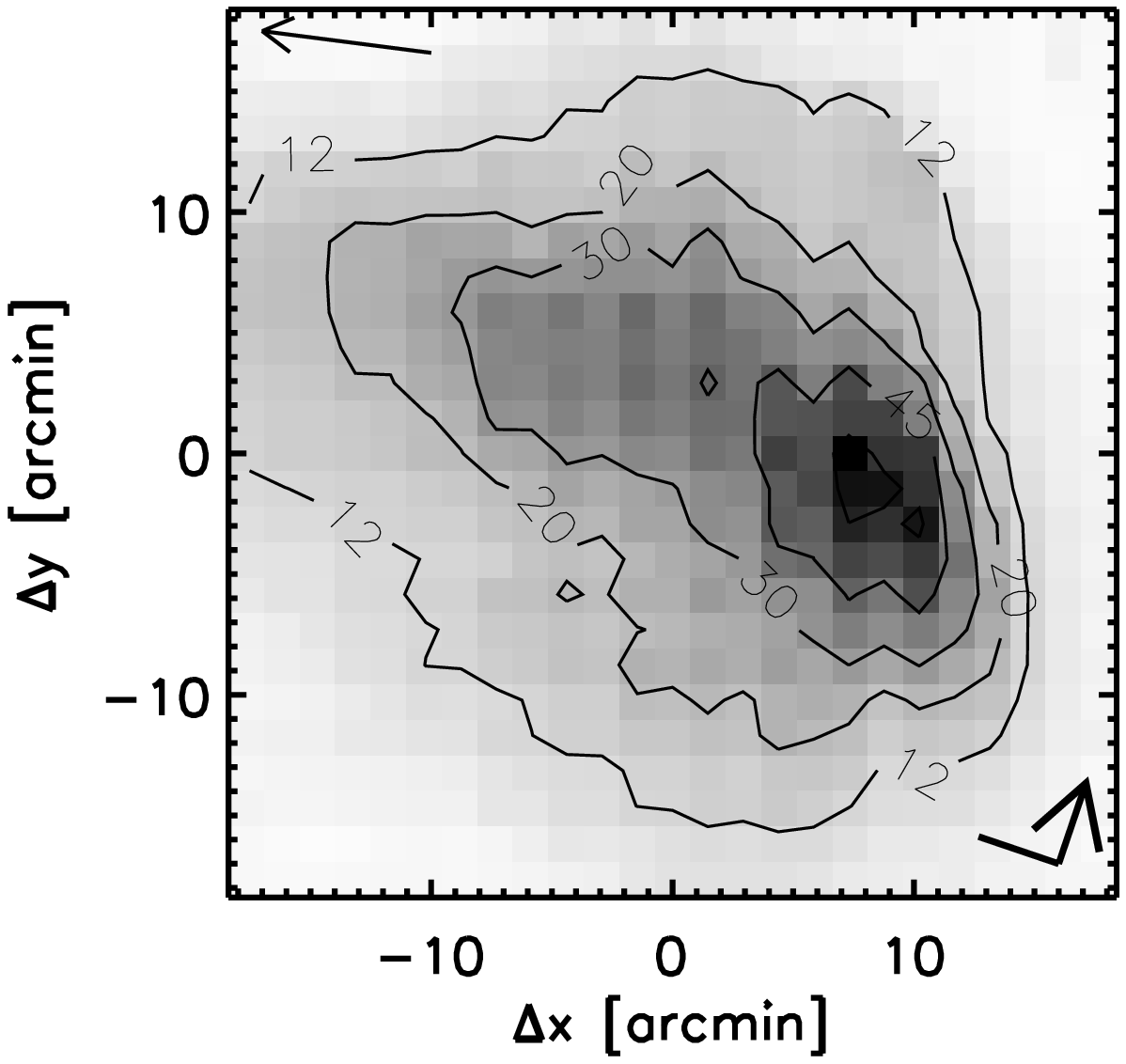,width=85mm,angle=0}}
\vspace*{0.3cm}
\caption{{\em Left}: {\em ISOPHOT} 100 $\mu$m emission smoothed to the angular resolution 
of the {\em ISOPHOT} 200 $\mu$m emission; {\em Right}: {\em ISOPHOT} 200 $\mu$m emission.
 Note the thin top-left arrow points to the Centaurus-Scorpius OB association. The NE axes are 
displayed on the bottom-right corner (arrow points to North).    
\label{fig:figure2}}
\end{figure*}

Fig. \ref{fig:figure1} and \ref{fig:figure2} show the {\em ISOPHOT\/} emission maps of
LDN 1780 for the stripe-like and the square regions, respectively. Maps in Fig. \ref{fig:figure2} 
were smoothed to the angular resolution of 200 $\mu$m {\em ISOPHOT} emission map 
(see Paper I). These maps have been zero-level calibrated following the procedure described in Sect. 2.4.

\subsection{Near-infrared extinction}

\begin{figure}
\vspace*{0.cm}\centerline{
\hspace*{1.25cm}\psfig{figure=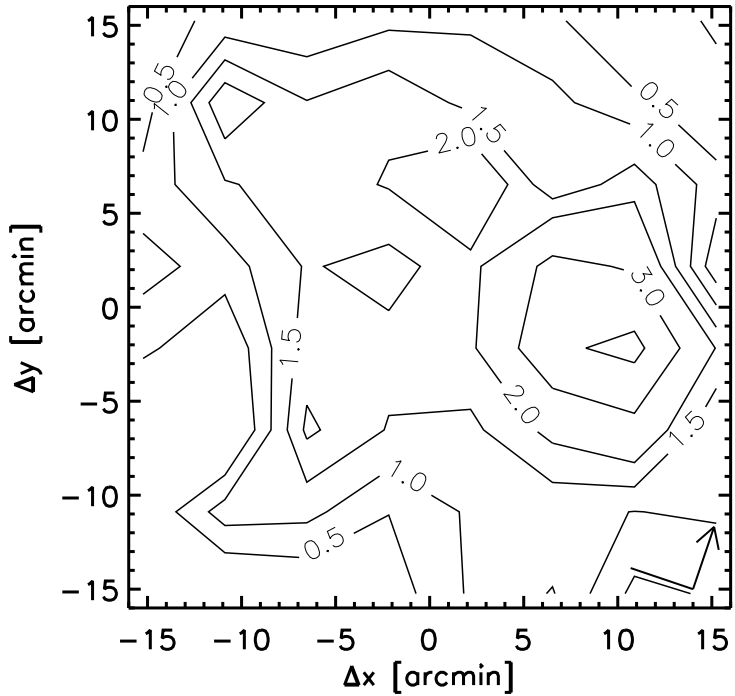,width=10cm,angle=0}}
\vspace*{0.cm}
\caption{Extinction map and contours in $A_V$ units. Pixel size of 
$4^{\prime}\times4^{\prime}$. Coordinates [0,0] correspond to 
$\alpha_{2000}$=15$^h$40$^m$11.6$^s$ and 
$\delta_{2000}$=-7$^{\rm o}$12$^{\prime}$13.5$^{\prime\prime}$.
\label{fig:figure3}}
\end{figure}

We built the extinction map of LDN 1780 following the method described in
Cambr\'esy et al. (2002) using the 2MASS $H-K_s$ colour excess of 3400 
stars in a $80^{\prime} \times 80^{\prime}$ field. The extinction is
obtained following the expression:

\begin{equation}
A_V = \left( \frac{A_H}{A_V} - \frac{A_{K_s}}{A_V} 
\right)^{-1} \times
(H-K_s) + \mathcal{Z}_{\rm col}
\end{equation}

where $(A_H/A_V - A_{K_s}/A_V)^{-1} = 15.87$ (Rieke \& Lebofsky 
1985) and $\mathcal{Z}_{\rm col}=-1.9$ mag. The zero point for the extinction
calibration, $\mathcal{Z}_{\rm col}$, corresponds to an intrinsic star
colour of $\overline{H-K_s}=0.12$ mag, measured outside the cloud.
Any diffuse extinction in the field is therefore removed by construction.
The extinction is obtained by taking the median value within adaptive
cells that contain a constant number of 3 stars. This produces a map
with a varying resolution, increasing with the star density.
The resulting map is convolved with an adaptive kernel to obtain the
final maps at the uniform resolutions of $4^{\prime}$ and $6^{\prime}$
(see Fig. \ref{fig:figure3}). Although one can na\"ively think this approach is equivalent
to directly use an uniform grid of $4'$ or $6'$, it is fundamentally
different. Because the ISM has small-scale structures,
variations of the extinction within a cell are expected. The direct
consequence of the sub-cell structures is that stars are not uniformly
distributed within a cell, but are preferentially detected in the lower
extinction region of a cell.
In the classical method with large cells, the few stars detected at higher
extinction do not contribute significantly to the colour median of a cell. The
weight of the low-extinction part of a cell is increased and the extinction
measurements are biased towards small values.
In the adaptive method with only 3 stars, cells have the minimum size and 
the final 4 or $6'$ resolution is obtained through a convolution.
The weight of the low extinction region does not dominate anymore since
these cells are smaller than at large extinction.
Using adaptive cells and a convolution give a weight to the individual
reddening measurements that is proportional to the surface of the voronoi
cell defined by 3 stars (the voronoi cell for a single star represents all
points equidistant of the central star to the surrounding objects).
This technique integrates both the star colour and density, which
optimises the extinction estimate by reducing the bias due to the
inhomogeneous star distribution within a cell (see Cambr\'esy et al.
2005 for more details).
The statistical uncertainties in the extinction maps are 0.4 mag and 0.5 mag
for the $6^{\prime}$ and $4^{\prime}$ resolution, respectively. At these
angular resolutions, the $6^{\prime}$ and $4^{\prime}$ extinction maps respectively
reach maxima of 3.9 mag and 4.4 mag in the central region of LDN 1780.

\subsection{Additional data}

In order to complement the {\em ISOPHOT} dataset of LDN 1780, the following 
additional data were included:
\begin{enumerate} 
\item {\em IRIS\/} maps (Miville-Desch\^{e}nes \& Lagache 2005) at 25, 60 
and 100 $\mu$m, with angular resolutions (noise levels in MJy sr$^{-1}$) of
3.8$^{\prime}$ (0.05), 4.0$^{\prime}$ (0.03) and
4.3$^{\prime}$ (0.06), respectively.
\item Carbon monoxide observations at 2.7 mm, (CO and $^{13}$CO) from LFHMIC95,
with an angular resolution of 2.7$^{\prime}$. In particular we have used the 
velocity-integrated (line integrated) emission maps of CO (W$_{12}$) and $^{13}$CO
(W$_{13}$).
\item H$\alpha$ surface brightness map, with an angular resolution of 
6$^{\prime}$, belonging to the H-alpha Full Sky Map (Finkbeiner 2003). The 
absolute calibration of the composite H$\alpha$ emission map is based on the
undersampled WHAM survey (1 deg spacing; 1 deg diameter top-hat beam). The error
in a 40$^{\prime}\times$40$^{\prime}$ H$\alpha$ emission map of LDN 1780 is $\sim$0.4 Rayleigh 
(1 Rayleigh=$\frac{10^6}{4\pi}$ $\gamma$ cm$^{-2}$ s$^{-1}$ sr$^{-1}$).
\end{enumerate}

\subsection{Data preparation}

We have applied different analysis approaches for the data analysis of the
stripe-like region and squared field data. 
In the remainder of this paper, we shall use the symbol $I_{\nu}$($\lambda$)
(in $\rmn{MJy\,sr^{-1}}$) to refer to the surface brightness, where
$\lambda$=25, 60, 70 and 100, 120 and 200~$\mu$m corresponds to the 
reference wavelengths of the {\em IRIS\/} and {\em ISOPHOT\/} filterbands. A 
distinction between {\em IRIS\/} and {\em ISOPHOT\/} data will be noticed if 
necessary. $I_\nu({\rmn H}\alpha)$ refers to the H$\alpha$ surface brightness. 
${\Delta}I_{\nu}$($\lambda$) refers to the error at the reference wavelength 
$\lambda$.

\subsubsection{Far-infrared emission correlations: stripe-like region}

Following the analysis of Paper I, we have derived colour variations from
the analysis of the correlations of $I_{\nu}(\lambda)$ ($\lambda<$200 $\mu$m)
with $I_{\nu}(200)$ corresponding to the stripe-like region. 
As described in Paper I, any colour variation will result in a deviation 
from linearity in the correlations. 
The method offers some advantages (see Paper I for a detailed discussion). Slopes of 
correlations are direct measurements of the colour ratios and are independent of 
any emission offset; we therefore did not apply any 
zero-level correction to the stripe-like region. 
In addition, slopes are much more accurate than the absolute surface brightness,
removing a significant amount of the random noise of individual pixels. 
In particular, when temperature  variations are within the uncertainties for 
a certain region, the use all independent points improves the accuracy and the 
method applied is optimal.

\begin{figure*}
\vspace*{-0.cm}\centerline{
\hspace*{0.1cm}\psfig{figure=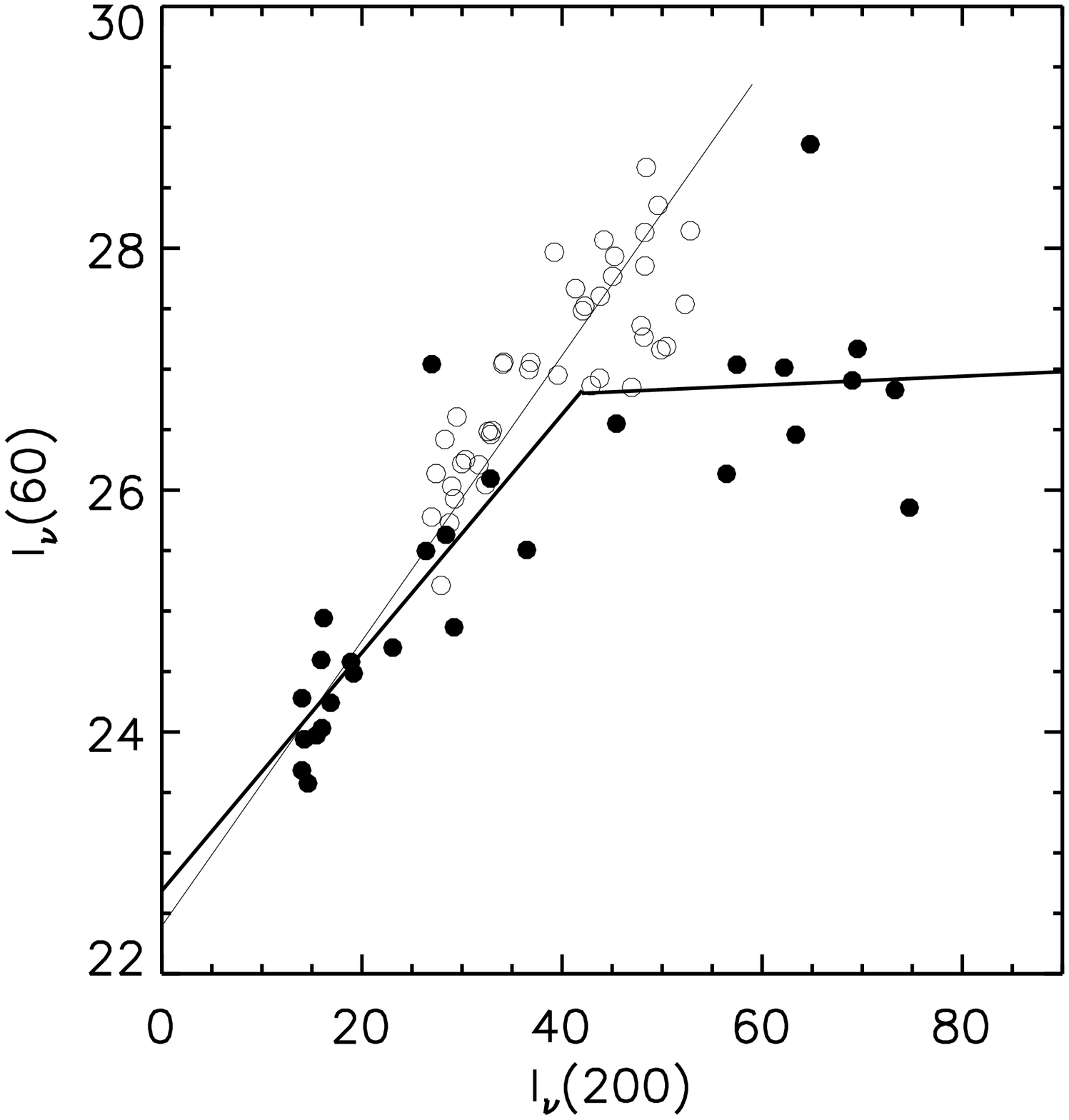,width=75mm,angle=0}
\hspace*{0.1cm}\psfig{figure=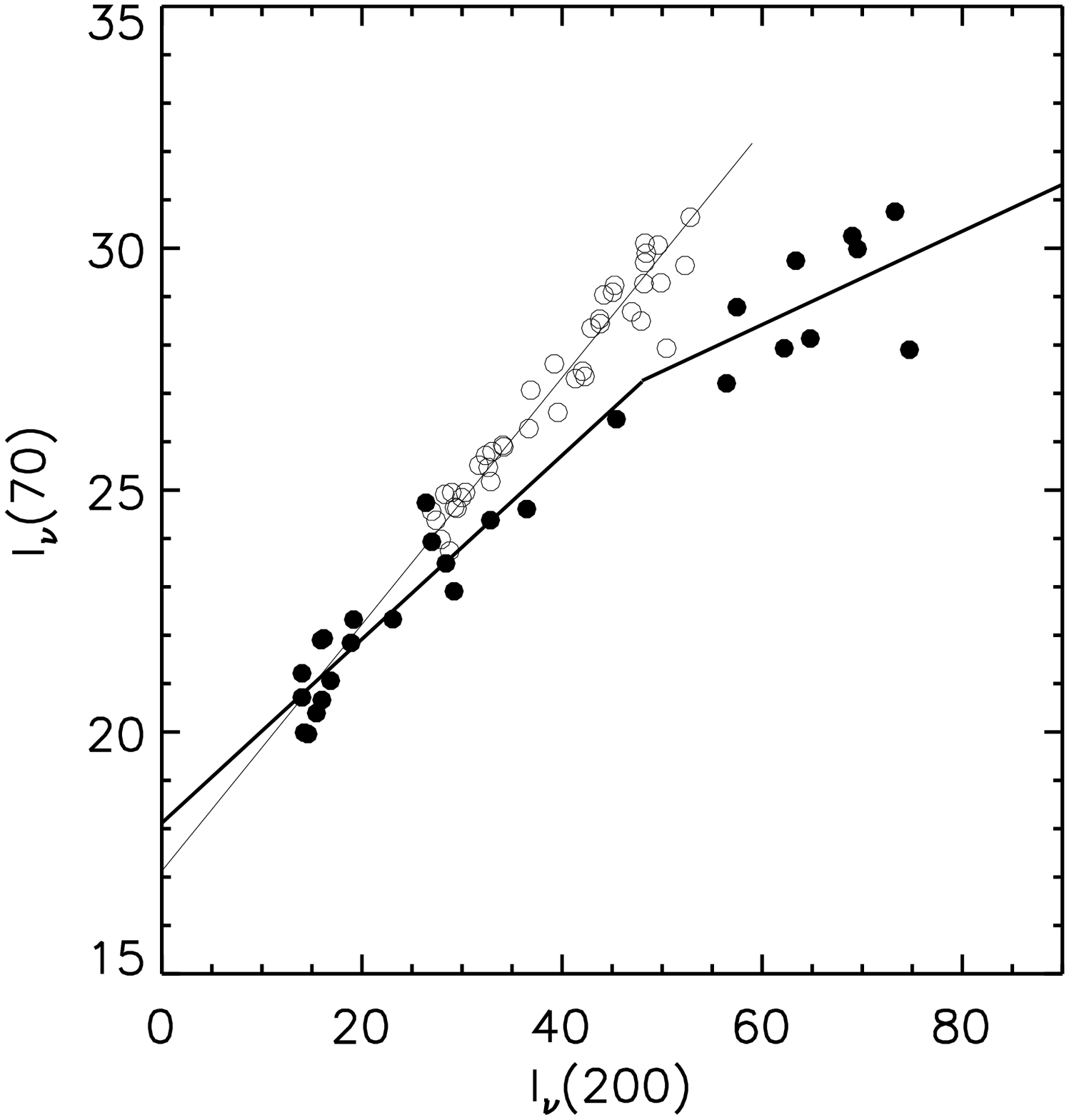,width=75mm,angle=0}}
\vspace*{0.cm}\centerline{
\hspace*{0.1cm}\psfig{figure=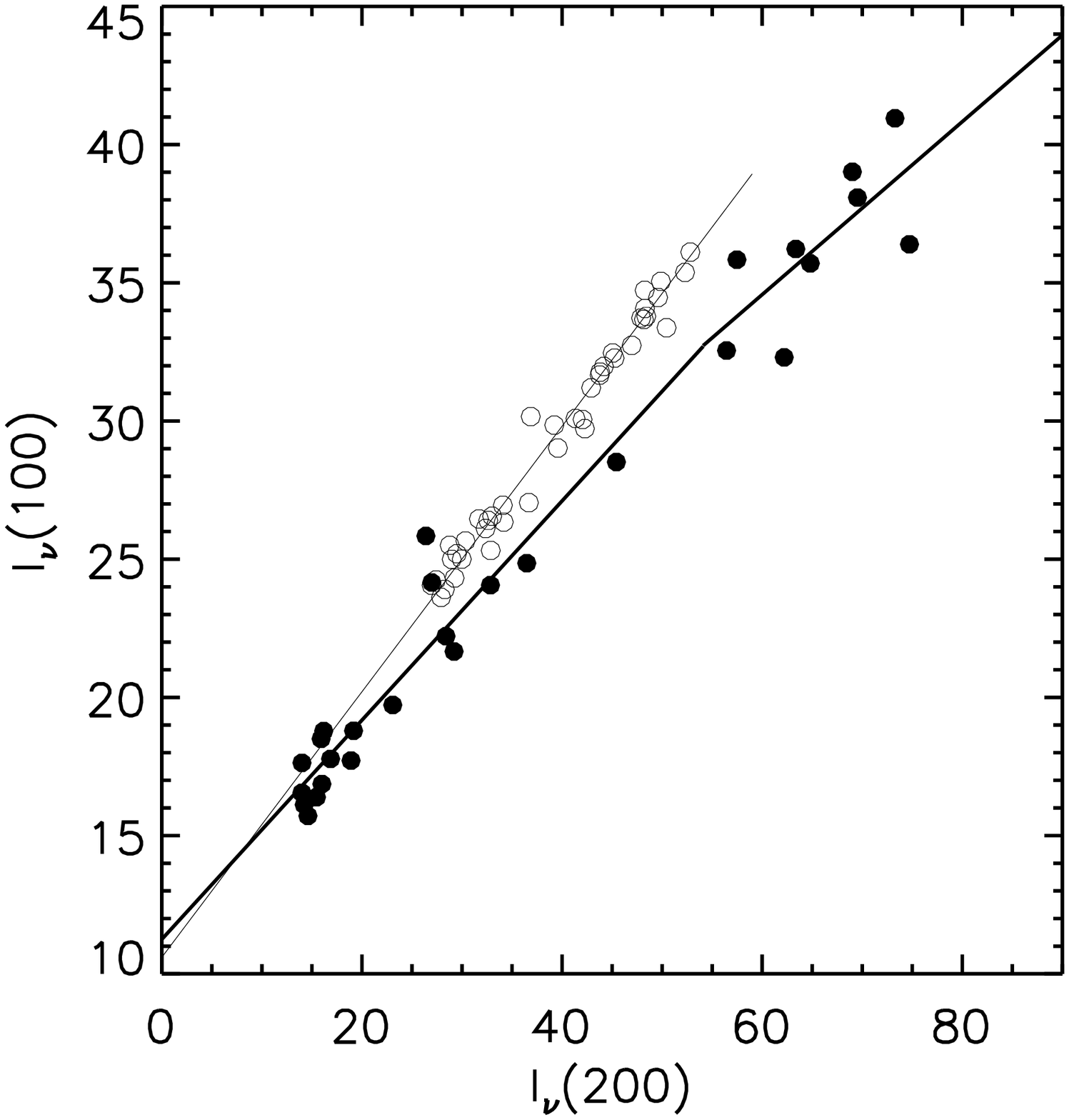,width=75mm,angle=0}
\hspace*{0.1cm}\psfig{figure=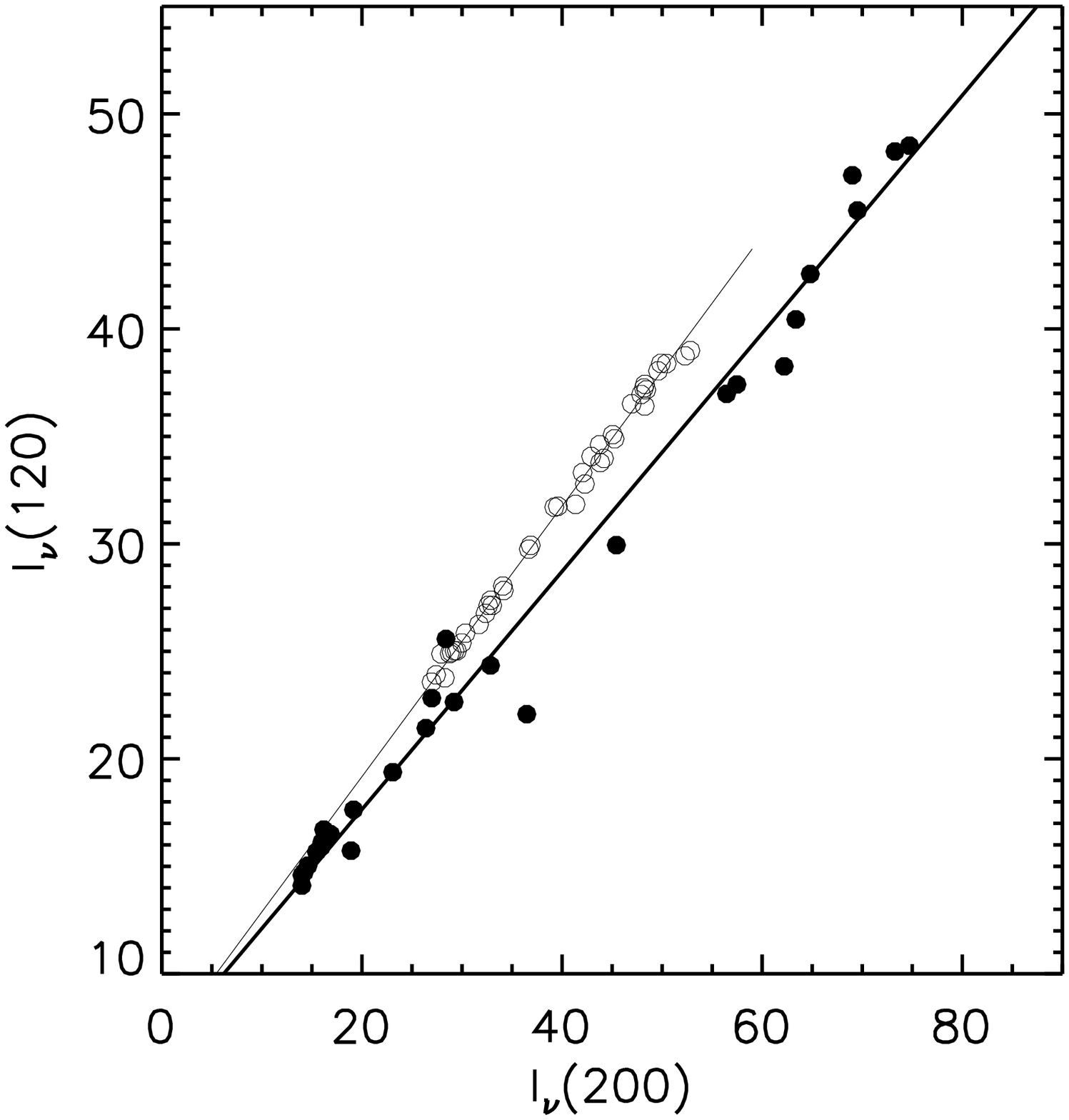,width=75mm,angle=0}}
\vspace*{0.cm}\centerline{
\hspace*{0.1cm}\psfig{figure=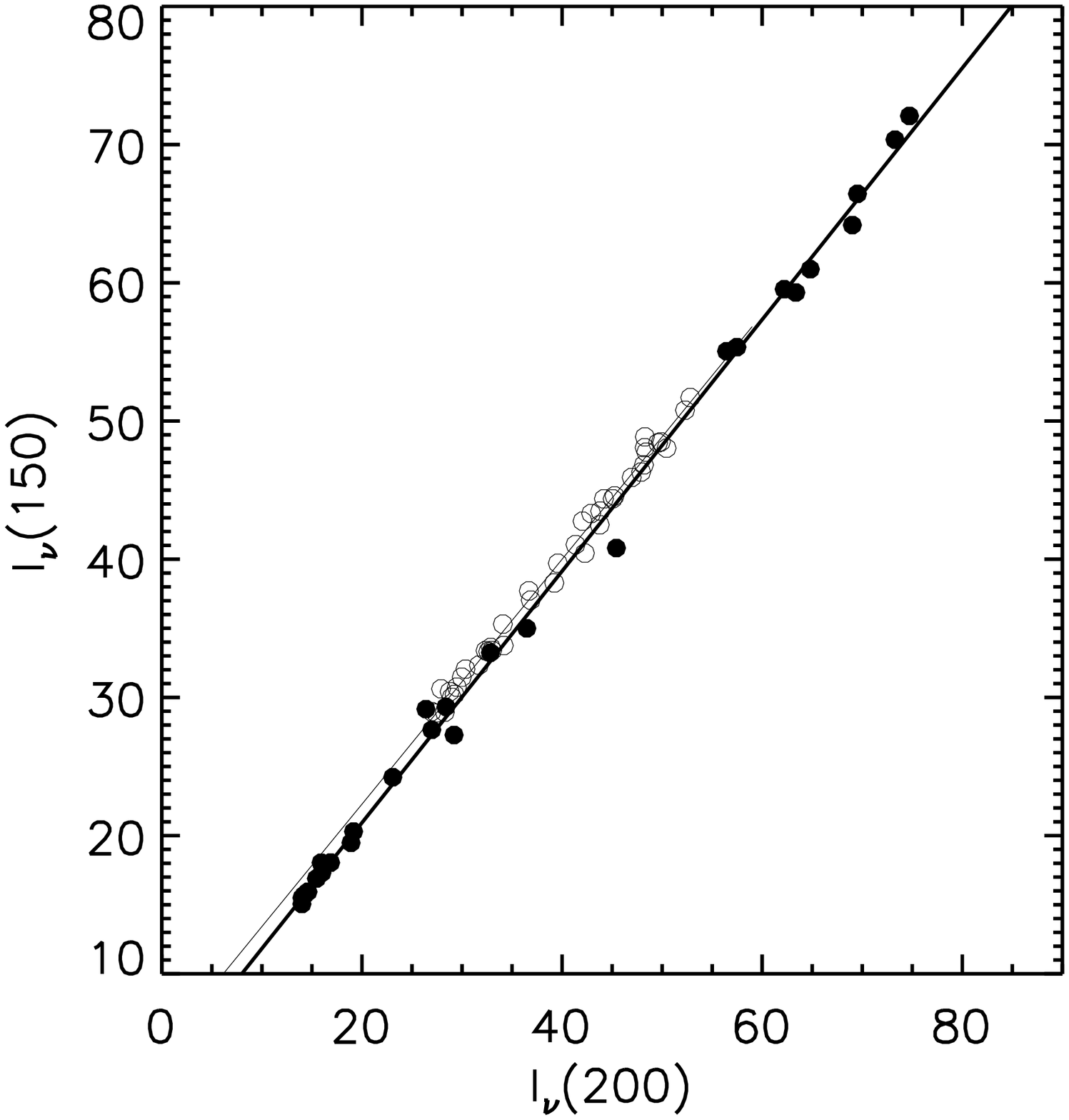,width=75mm,angle=0}}
\vspace*{0.3cm}
\caption{Surface brightness $I_{\nu}$($\lambda$) versus $I_{\nu}$(200)
for wavelengths of 60, 70, 100, 120 and 150 $\mu$m. White and black circles
respectively correspond to the illuminated region and shadowed region of LDN 1780.
Straight lines that fit the data points are overplotted (see Sects. 2.4.1 and 3.1.1).}
\label{fig:figure4}
\end{figure*}

All the stripe-like region data were obtained using {\em ISOPHOT}. 
Data corresponding to the C100 camera were smoothed to the angular resolution of the 2$\times$2 pixel C200 camera.
Fig. \ref{fig:figure4} shows the correlations for the emissions at 60, 70 and 100, 120 versus
emission at 200 $\mu$m. Correlations for two separated regions of LDN 1780 have
been considered (see Fig. \ref{fig:figure1}, {\em bottom}): a region in 
the shadow part of the cloud and a similar-in-size region
in the illuminated side of the cloud, which faces the Galactic plane and the OB association. 
The instrumental noises of these emissions have been determined 
from the pixel to pixel response variations derived from the FCS measurements 
(see Table~\ref{isophot:noise}).

\subsubsection{Map homogenisation: squared field}

The maps of {\em IRIS}, {\em ISOPHOT}, the carbon monoxide lines as well as 
the H$_\alpha$ surface brightness map for the squared field 
have different angular resolutions.
For a proper comparison the maps were smoothed to the worst angular
resolution\footnote{We excluded the HI excess emission map for this 
comparison.}. For the analysis of far-infrared, extinction and molecular
data, maps were smoothed to the angular resolution of the 100 $\mu$m 
{\em IRIS} map of 4.3$^{\prime}$ (corresponding to 0.14 pc for a distance of 
110 pc). For a comparison with the H$\alpha$ surface brightness with 
FWHM$\approx$6$^{\prime}$, the maps were smoothed to mimic that angular
resolution.

\begin{figure*}
\vspace*{-.75cm}\centerline{
\hspace*{0.1cm}\psfig{figure=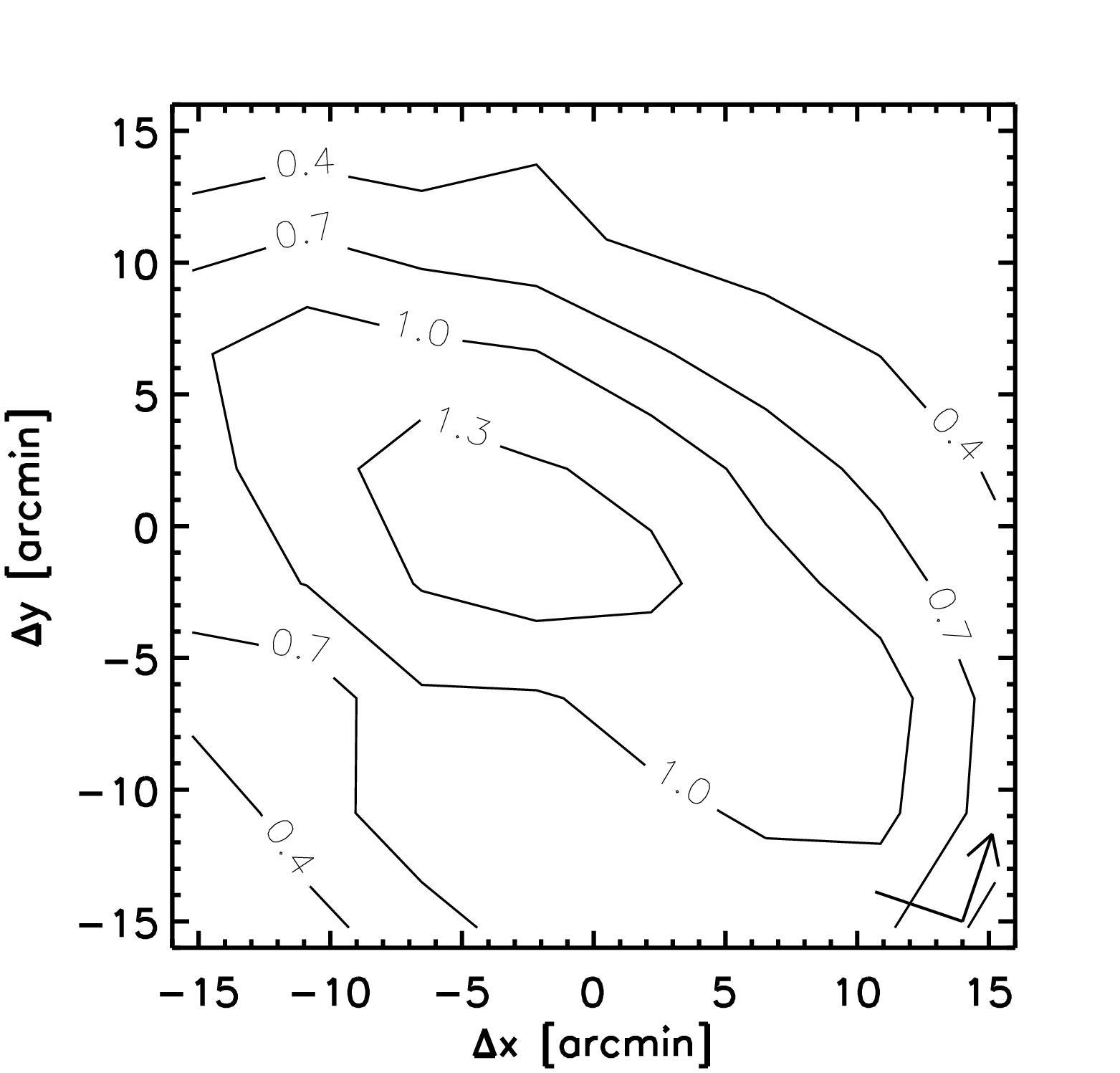,width=77mm,angle=0}
\hspace*{0.1cm}\psfig{figure=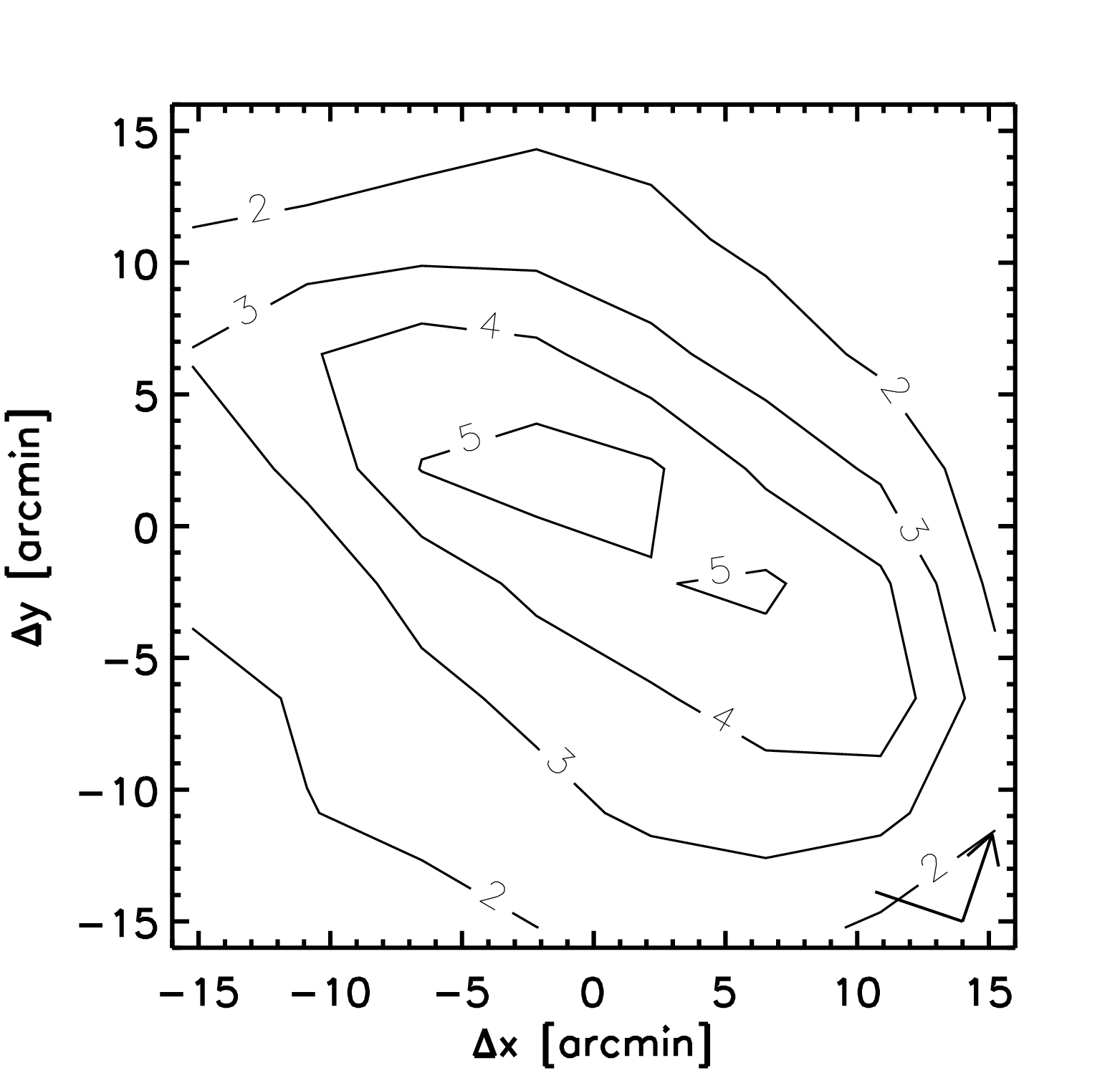,width=77mm,angle=0}}
\vspace*{-.5cm}\centerline{
\hspace*{0.1cm}\psfig{figure=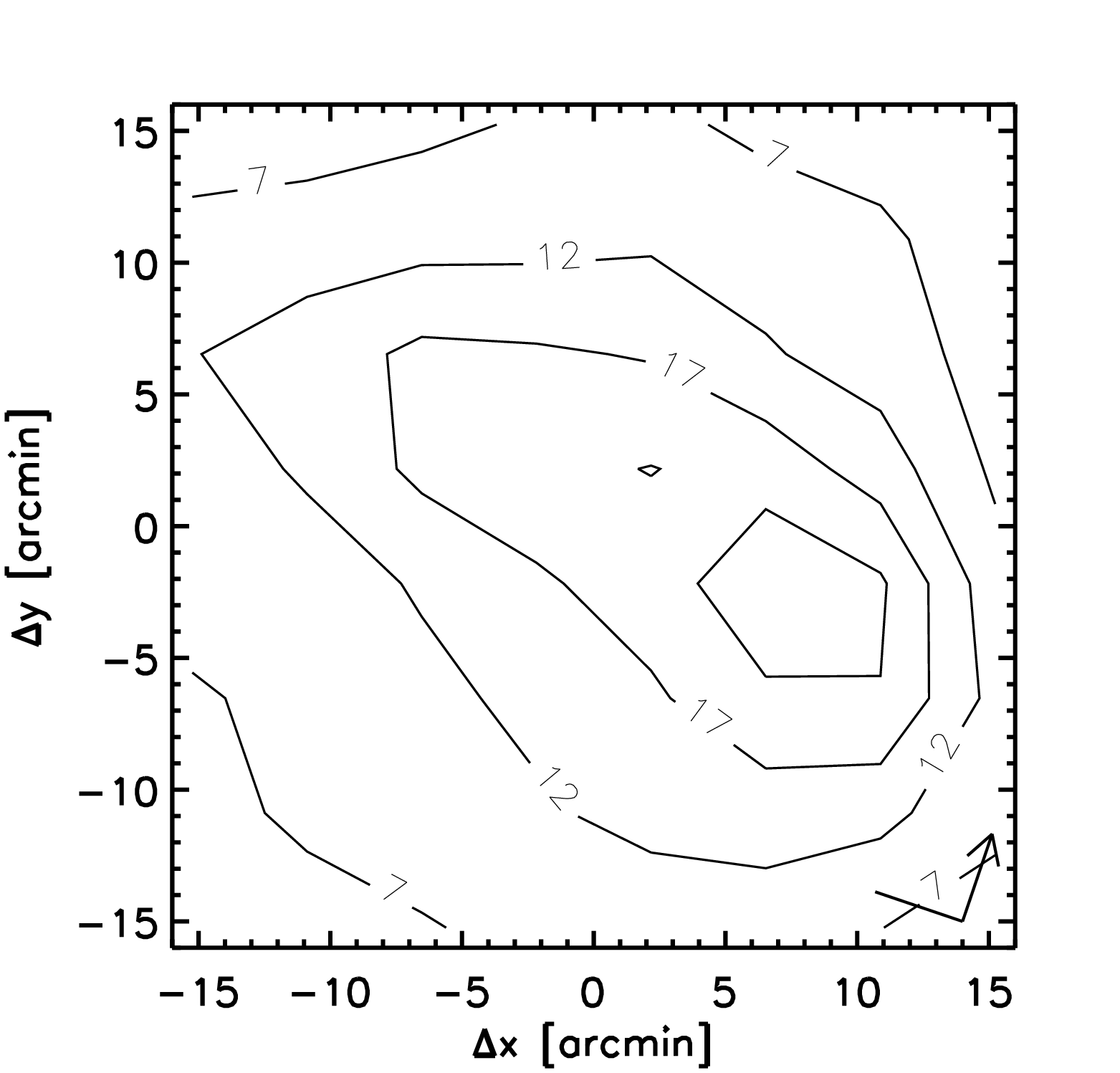,width=77mm,angle=0}
\hspace*{0.1cm}\psfig{figure=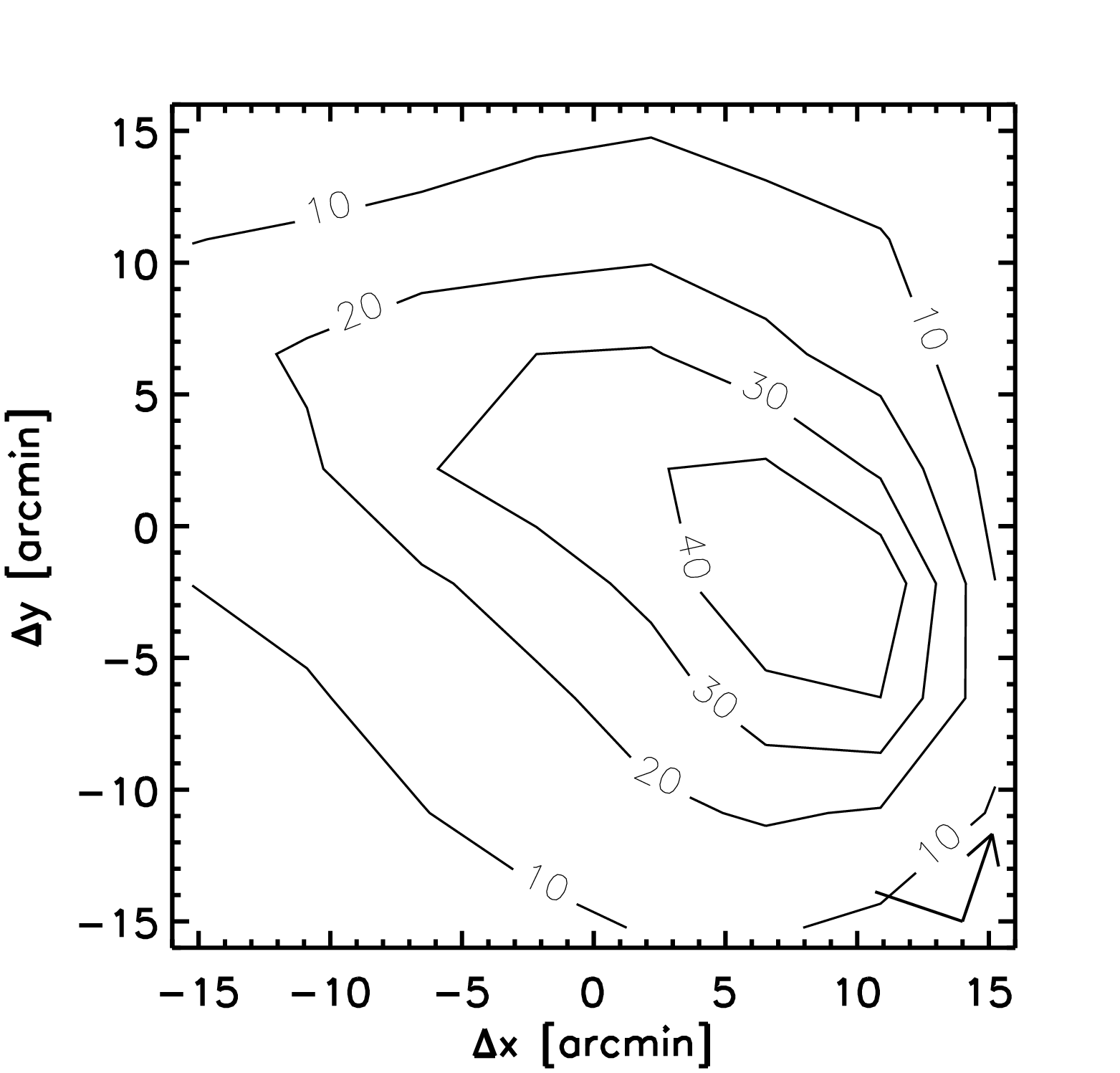,width=77mm,angle=0}}
\vspace*{-.5cm}\centerline{
\hspace*{0.1cm}\psfig{figure=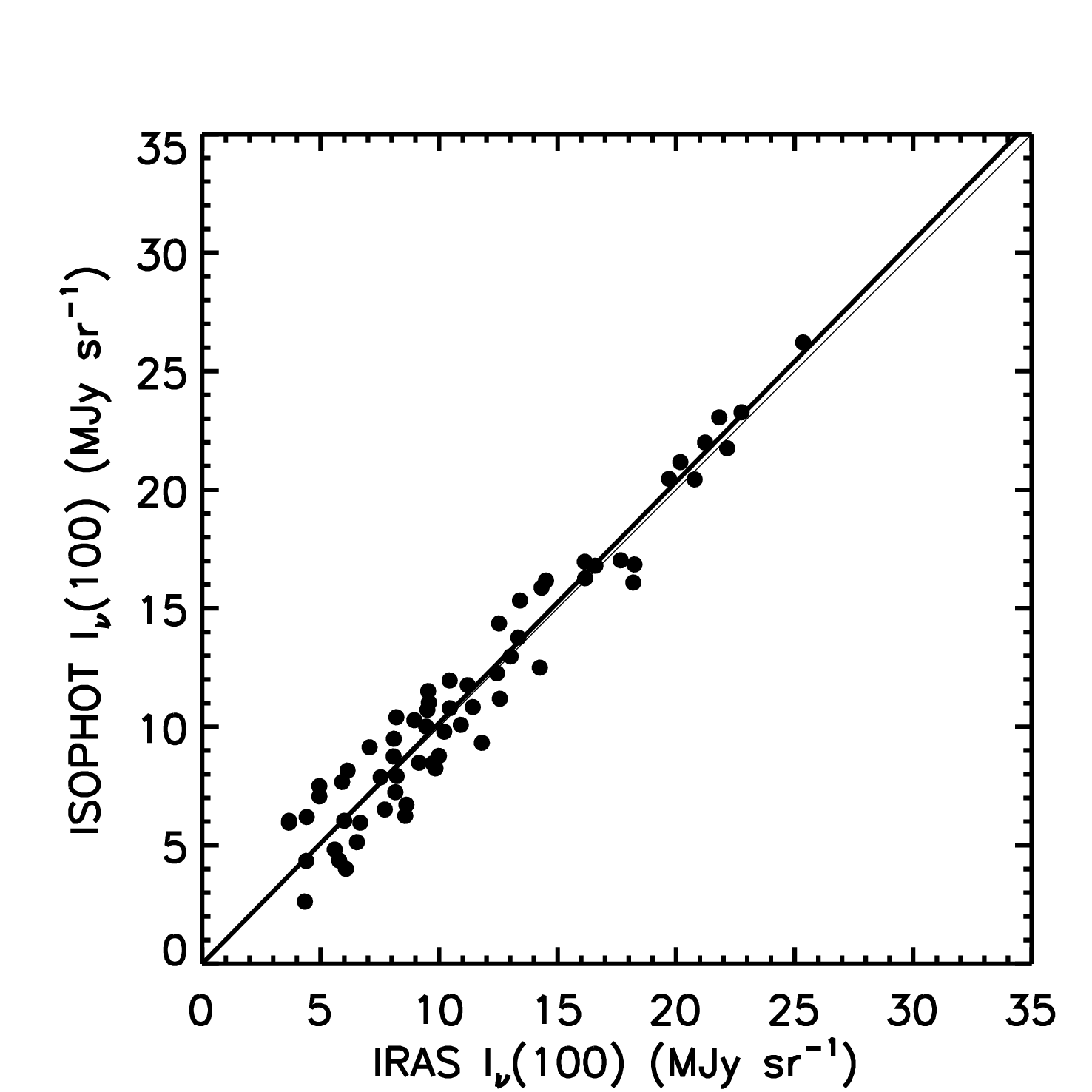,width=77mm,angle=0}
\hspace*{0.1cm}\psfig{figure=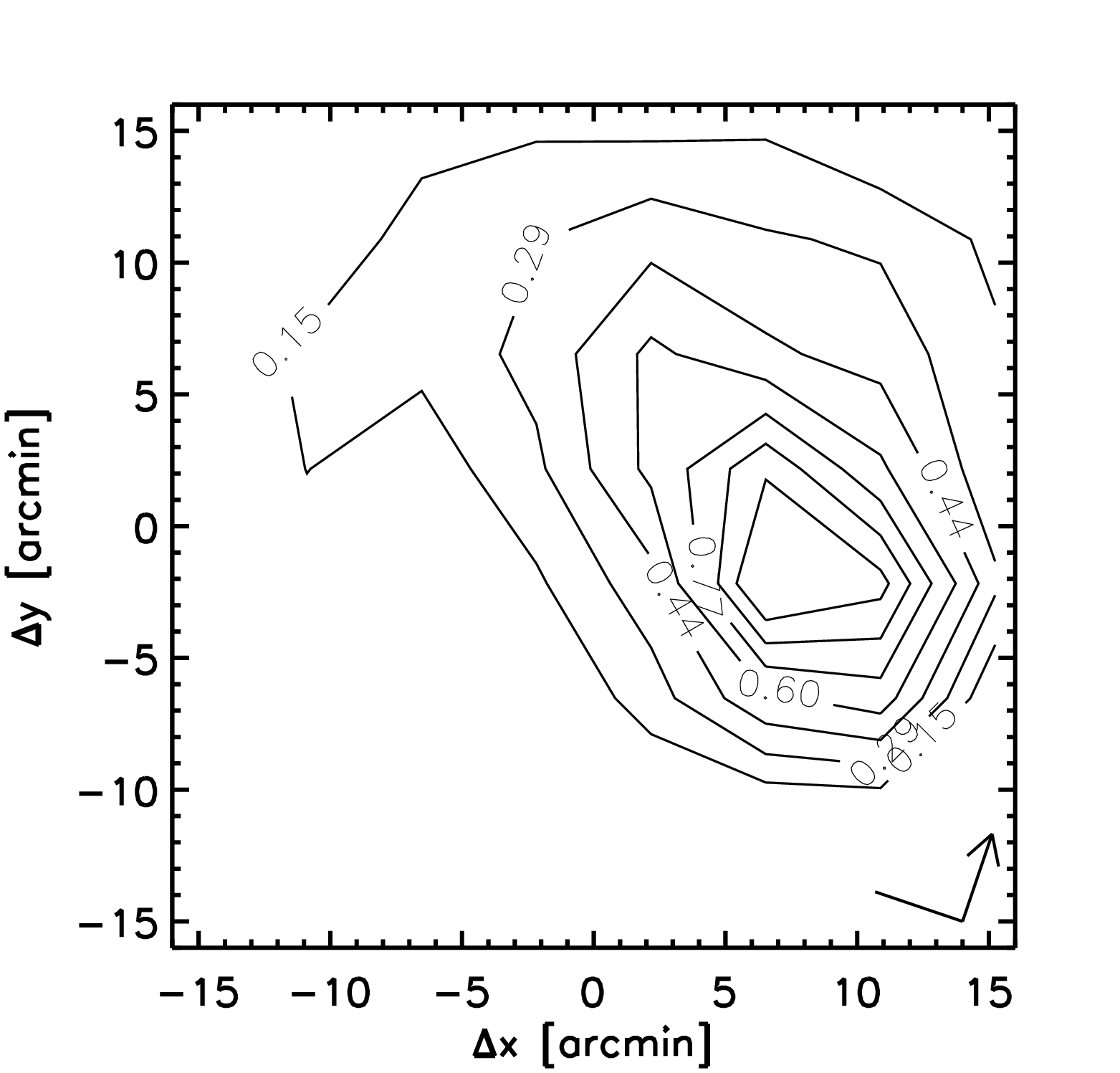,width=77mm,angle=0}}
\vspace*{0.3cm}
\caption{{\em Top-left:} {\em IRIS} 25 $\mu$m emission map; 
{\em Top-right:} {\em IRIS} 60 $\mu$m emission map; 
{\em Middle-left:} {\em IRIS} 100 $\mu$m emission map; 
{\em Middle-right:} {\em ISOPHOT} 200 $\mu$m emission map.
All maps with angular resolution of 4.3$^{\prime}$. 
The NE axes (arrow points to North) are displayed on the 
bottom-right corner of each map. The thin and thick lines respectively
correspond to the 1:1 and observed relations; 
{\em Bottom-left:} {\em ISOPHOT} 100 $\mu$m emission versus {\em IRIS} 100 $\mu$m emission.
Note emission are not colour-corrected. The thin and thick lines respectively correspond to the 1:1 and observed relations; 
{\em Bottom-right:} $^{13}$CO(J=1-0) line integrated. Units of K Km s$^{-1}$.
}
\label{fig:figure5}
\end{figure*}

In Fig. \ref{fig:figure5} the 25, 60 and 100 $\mu$m {\em IRIS} maps (see 
Fig. \ref{fig:figure5}, {\em top},
{\em middle-left}) and 200 $\mu$m {\em ISOPHOT} map (see Fig. 
\ref{fig:figure5}, {\em middle-right})
at the angular resolution of 4.3$^{\prime}$ are shown.

The median instrumental noise in $I_{\nu}(25)$, $I_{\nu}(60)$,
$I_{\nu}(100)$ reported by Miville-Desch\^enes \& Lagache (2005) is
0.05, 0.03 and 0.06 MJy sr$^{-1}$, respectively. For {\em ISOPHOT}, the pixel
to pixel response variations derived from the FCS measurements (see Sect.
2.1) have been used to determine the instrumental noise. Table~\ref{isophot:noise} 
shows the values of the instrumental noise at the
different wavelengths for the observations. Data were
smoothed to the $4.3^{\prime}$ angular resolution. When
data from different instruments are combined we have also considered the
relative errors between different filterbands. In Paper II we used a 5\%
between {\em ISOPHOT} filterbands and a pessimistic 30\% for the relative
error between {\em IRIS} 100 $\mu$m and the {\em ISOPHOT} 200 $\mu$m. Here, we have
correlated the {\em IRIS} and smoothed 4.3$^{\prime}$ {\em ISOPHOT} 100
$\mu$m maps (see Fig. \ref{fig:figure5}, {\em bottom-left}) to quantify this 
relative error in LDN 1780. We found an excellent correlation (Pearson correlation 
coefficient of PCC=0.97) between these maps and a relative error of only 1.6\%. 
This value has been used as
reference to derive the relative errors between {\em IRIS} and {\em ISOPHOT}
filterbands assuming a 5\% between different filterbands of the same
instrument.

\begin{table*}\begin{center}\begin{footnotesize}
\caption{{\em ISOPHOT\/} instrumental noise. \label{isophot:noise}}
\begin{tabular}{ccccccc}
\hline
${\Delta}I_\nu(60)$ & ${\Delta}I_\nu(70)$ & ${\Delta}I_\nu(100)$ & ${\Delta}I_\nu(150)$ & ${\Delta}I_\nu(200)$ & Remark \\
 $[MJy\ sr^{-1}]$ & $[MJy\ sr^{-1}]$ & $[MJy\ sr^{-1}]$ & $[MJy\ sr^{-1}]$ & $[MJy\ sr^{-1}]$ & \\ \hline
 0.4 & 0.35 & 0.3 & 0.55 & 0.5 & stripe at $1.5^{\prime}$ \\
 - & - & 0.45 & - & 0.76 & square at $1.5^{\prime}$ \\
 - & - & 0.16 & - & 0.27 & square at $4.3^{\prime}$ \\
\hline
\end{tabular}\end{footnotesize}\end{center}
\end{table*}

\subsubsection{Separation of the warm and cold components}

The emission maps of the squared field in LDN 1780 at 25, 60, 100 and 200 $\mu$m show
distinct morphologies that likely indicate variations in the abundance
distribution across the cloud. Note, for instance, the different positions of
the emissions peaks at different wavelengths in Fig. \ref{fig:figure5}. See also
Fig. \ref{fig:figure1} for intermediate wavelengths.

In order to distinguish between the different large dust grain components in 
LDN 1780 we have performed a first attempt to 
obtain the emission of the cold component at 100 $\mu$m (see Laureijs, Clark \& Prusti 
1991; Abergel et al. 1994), and at 200 $\mu$m (see Paper II) using the equation:

\begin{equation}
{\Delta}I_\nu(\lambda)(x,y)=I_{\nu}(\lambda)(x,y)-I_{\nu}(60)(x,y)/\Theta(60,\lambda),
\end{equation}

where ${\Theta}(60,\lambda)$ is the average colour $I_\nu(60)/I_\nu(\lambda)$
of LDN 1780. We have determined ${\Theta}(60,100)$=0.27
from the {\em IRIS} maps, which is the same value found by LFHMIC95
from previously realised {\em IRAS} data. Our value of ${\Theta}(60,\lambda)$
was determined using values of $I_\nu(100)$ below 90\% of the peak of $I_\nu(100)$. 
We found ${\Theta}(60,200)$=0.13 using values of $I_\nu(200)$ below 90\% of the maximum.
The correlation between ${\Delta}I_\nu(100)$ and $I_\nu(200)$ has a PCC of 0.7.

The previous approach has been traditionally used to separate the
BGs that are associated to the molecular regions and are free 
from VSG emission from the warm diffuse ISM. Although they were originally
called {\em warm} and {\em cold}, these components separate two grain
populations rather than two temperatures. 
In addition recent results obtained at longer wavelengths using far-infrared
and submillimeter instruments indicate the presence of two {\em temperature}
components of BGs, with temperatures of $\approx$16-22 \,K
and $\sim$12\,K, respectively, and different emissivities 
(Cambr\'esy et al. 2001, Papers I and II).
The designation of {\em warm} and {\em cold} components is still used by the 
community to refer to this separation.

A {\em second approach} has been used to carry out the dust component
separation. Before introducing this approach, we give a brief scientific justification for
its use, which will be discussed in more detail in Sect. 3.6.
Assuming a single BG component, the observed mean
ratio $I_{\nu}(100)/I_{\nu}(200)$ would indicate a temperature of 
$\sim$17\,K in the cloud, which is relatively high compared with other
translucent clouds (see Paper I). Also, this temperature is significantly lower than the 
temperature of $\sim$13\,K derived from the ${\Delta}I_{\nu}(100)/I_{\nu}(200)$ ratio using a modified black body with a power-law emissivity $\beta$=2.
This discrepancy in temperature can be interpreted as due to the presence of two BG component
instead of a single one.
This could explain other findings in the cloud.
It is observed that the 25 $\mu$m emission resembles properties
of both the 12 and 60 $\mu$m emission maps.
 In addition, the average colour $\Theta(60,100)$=0.27
is relatively high with respect to the standard value in the ISM of
$\Theta(60,100)$=0.21. LFHMIC95 suggested that the high ratio $\Theta(60,100)$
in LDN 1780 might be due to a high contribution of VSGs, but could be
alternatively interpreted as due to the presence of a warm component of BGs.
The enhanced stellar radiation field in LDN 1780 (LFHMIC95, THML95) due to
the Galactic plane and a close OB association could produce
that the warm and cold components present relatively high temperatures compared 
with those found in Cambr\'esy et al. (2001), Papers I and II. A significant
contribution to the emission at 60 $\mu$m from warm BGs is therefore expected. 
The 25 $\mu$m emission, which presents a high
S/N ratio, becomes the suitable template of the VSG emission.

Hereinafter, when referring to the emission of the warm and cold components we 
respectively use the superscripts $w$ and $c$. We determined the emission of 
the cold component at an intermediate wavelength (60, 100 and 200 $\mu$m) from 
the following equation:

\begin{equation}
I^c_\nu(\lambda)(x,y)=I_{\nu}(\lambda)(x,y)-c(\lambda,25) \times 
I_{\nu}(25)(x,y),
\end{equation}

where $c(\lambda,25)$ is a constant value derived from the correlation
$I_{\nu}(\lambda)$-$I_{\nu}(25)$. It corresponds to the slope of this
correlation, i.e. $I_{\nu}(\lambda)$/$I_{\nu}(25)$, for values of
$I_{\nu}(\lambda)$ below the 90\% of its maximum. This is performed to reject out 
possible outliers in the emission maxima surroundings.

In order to determine the emission of the warm component at wavelengths
$\lambda=$25, 60 and 100 $\mu$m we applied the equation:

\begin{equation}
I^{w}_\nu(\lambda)(x,y)=I_{\nu}(\lambda)(x,y)-w(\lambda,200){\times}I^c_\nu(200)(x,y),
\end{equation}

where $w(\lambda,200)$ is the ratio $I_{\nu}(\lambda)$/$I^c_{\nu}(200)$. 
We have simply used $I_\nu(200)$ instead of $I^c_\nu(200)$ when determining 
$I^{w}_\nu(25)$.

Note this second approach provides emission maps at wavelengths 60, 100 and 200 $\mu$m
for the cold component. The 100 and 200 $\mu$m emission maps will be used to derive colour
temperature and optical depth maps in the Sections 3.2.2 and 3.3, respectively.
For the warm component emission maps at 25, 60 and 100 $\mu$m are obtained.
The latter two maps have been used to determine the colour temperature and optical depth
of the warm component (see sections 3.2.3 and 3.3, respectively).

\subsubsection{Zero level calibration}

We have fixed to zero the minimum value of $I^c_\nu(100)$ in the field of view of
{\em ISOPHOT}. The region is representative of the local background for LDN
1780. Then, the pixel-pixel correlations of $I^c_\nu(60)$ with $I^c_\nu(100)$
and $I^c_\nu(200)$ with $I^c_\nu(100)$ where used to derive the offsets of
the emissions at 60 and 100 $\mu$m with respect the emission at 200 $\mu$m. 
We have estimated an error of ${\Delta}I^c_\nu(200)$=1 MJy 
sr$^{-1}$ due to the zero level uncertainty.

For the warm component we have fixed the minimum of $I^w_\nu(100)$ to zero.
Then, the zero levels of the emissions at 25, and 60 $\mu$m and their
corresponding errors were determined from the
pixel-pixel correlations with the corrected $I^w_\nu(100)$. 
We have computed an error at the reference 
wavelength of 100 $\mu$m of ${\Delta}I^w_\nu(100)$=0.15 MJy sr$^{-1}$ due to 
the zero level uncertainty.

We were consistent with these definitions for the zero level calibration when
applying the method described in the previous section, i.e, when removing the
emissions at 25 and 200 $\mu$m emissions. We used the zero-level errors 
together with the relative errors between filterbands and the errors due to the 
instrumental noise when computing the temperature errors and the optical depth 
errors in the next sections.

\section{Physical properties of the warm and cold components}

\subsection{Far-infrared emission}

\subsubsection{Striped region}

We have determined the colour variations along the stripe-like region, in the illuminated
and shadowed sides (see Fig. \ref{fig:figure1}, {\em bottom}). For both sides, 
the correlations $I_\nu(\lambda)-I_\nu(200)$ shown in Fig. \ref{fig:figure4} present a linear
behave within the scatter at wavelengths of 120 and 150 $\mu$m. This is described as
{\em unimodal} behave in Paper I. At shorter wavelengths it is observed that each
correlation is {\em bimodal} for the shadowed region: the colour variation is
modelled by two connected linear ramps with a wavelength dependent turnover point. 
The observed flattening in the emissions towards higher
200 $\mu$m emission indicates a decrease in the colour ratio at higher column densities.

The slopes of the correlations for the unimodal and bimodal cases are determined from
a least-squares linear fit. The regression fits are overplotted on the correlations
in Fig. \ref{fig:figure4} and the slopes are listed in Table \ref{isophot:slopes}. $Slope_1$ and $Slope_2$ stand for the slope values of the correlations before and after the turnover emission point.

\begin{table}\begin{center}\begin{footnotesize}
\caption{Slopes in the regressions of the striped region observed using {\em ISOPHOT}. 
Uncertainties in the slopes correspond to statistical variations. \label{isophot:slopes}}
\begin{tabular}{lccc}
\hline
 Region      & $\lambda_{ref}$ & Slope$_1$       & Slope$_2$     \\
             & ($\mu$m)        &                 &               \\ \hline
 Shadowed    & 60              & 0.10$\pm$0.02   & 0.00$\pm$0.02 \\
             & 70              & 0.19$\pm$0.01   & 0.10$\pm$0.06 \\
             & 100             & 0.40$\pm$0.02   & 0.3$\pm$0.1   \\
             & 120             & 0.55$\pm$0.01   & -             \\
             & 150             & 0.910$\pm$0.007 & -             \\
             & 200             & 1               &           \\ \hline
 Illuminated & 60              & 0.118$\pm$0.009 & -         \\
             & 70              & 0.255$\pm$0.009 & -         \\
             & 100             & 0.48$\pm$0.01   & -         \\
             & 120             & 0.629$\pm$0.007 & -         \\
             & 150             & 0.89$\pm$0.01   & -         \\
             & 200             & 1               & -         \\
\hline
\end{tabular}\end{footnotesize}\end{center}
\end{table}

\subsubsection{Squared region}

\begin{figure*}
\vspace*{-0.3cm}\centerline{
\hspace*{0.1cm}\psfig{figure=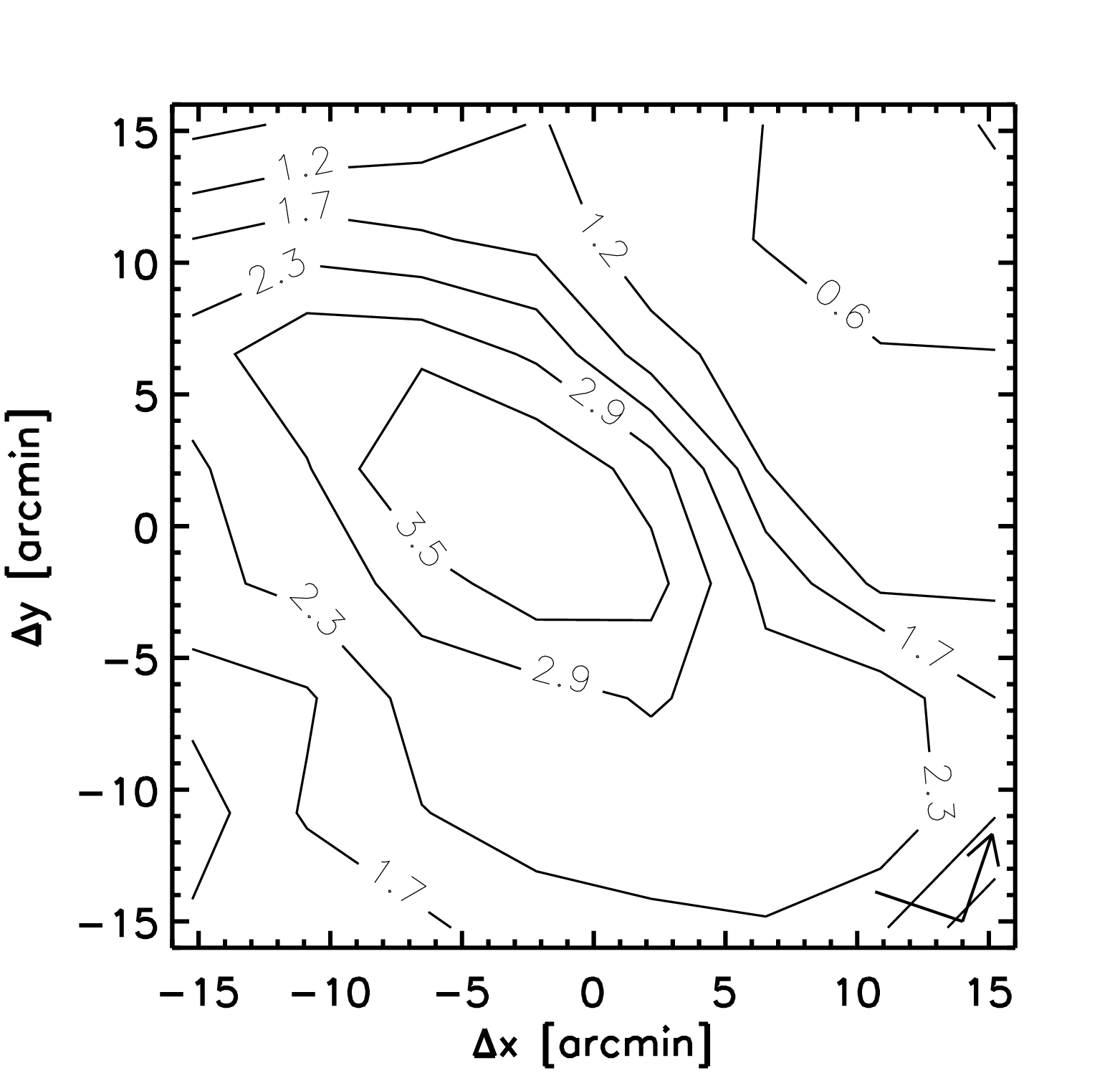,width=75mm,angle=0}
\hspace*{0.1cm}\psfig{figure=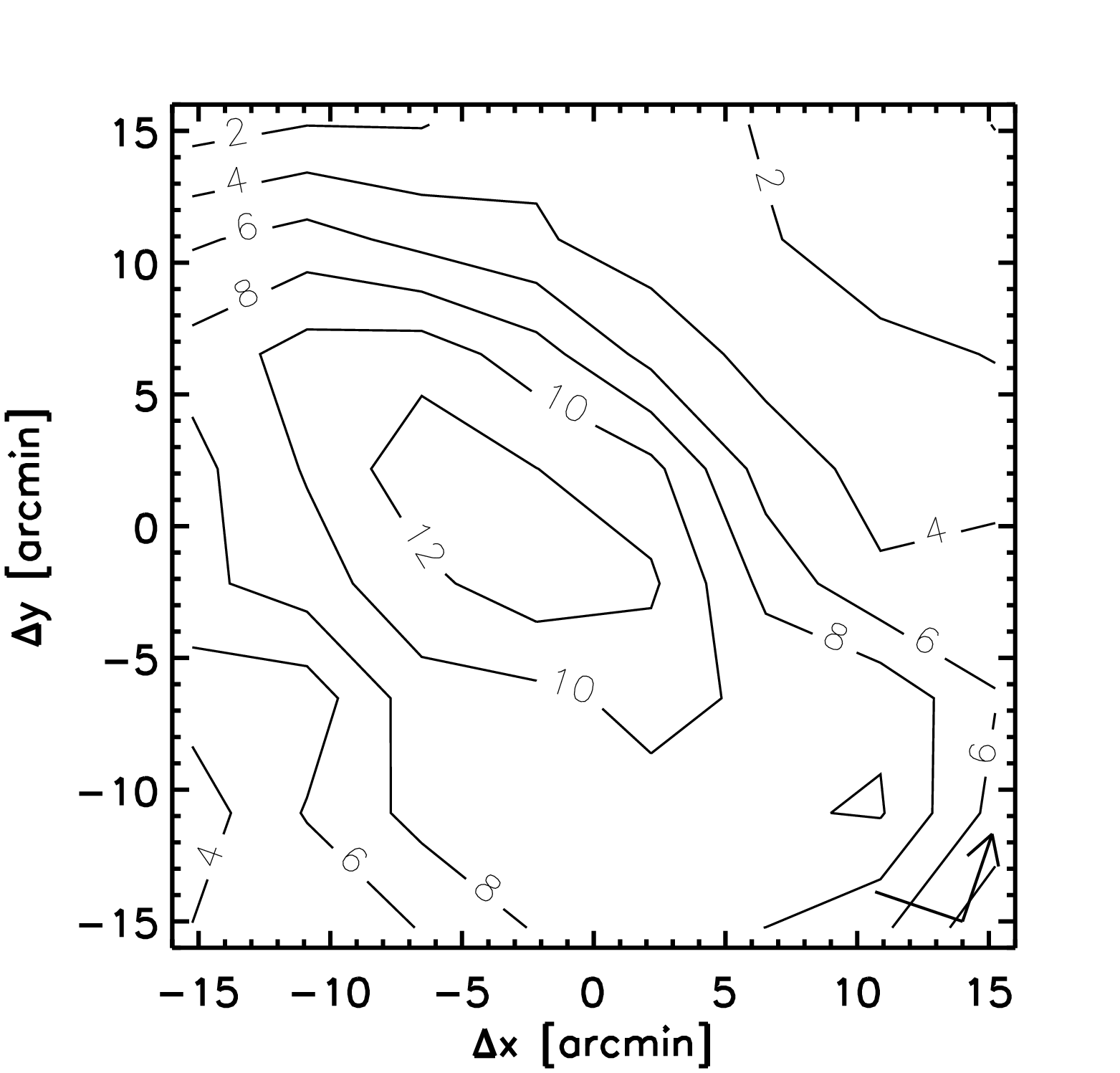,width=75mm,angle=0}}
\vspace*{-0.3cm}\centerline{
\hspace*{0.1cm}\psfig{figure=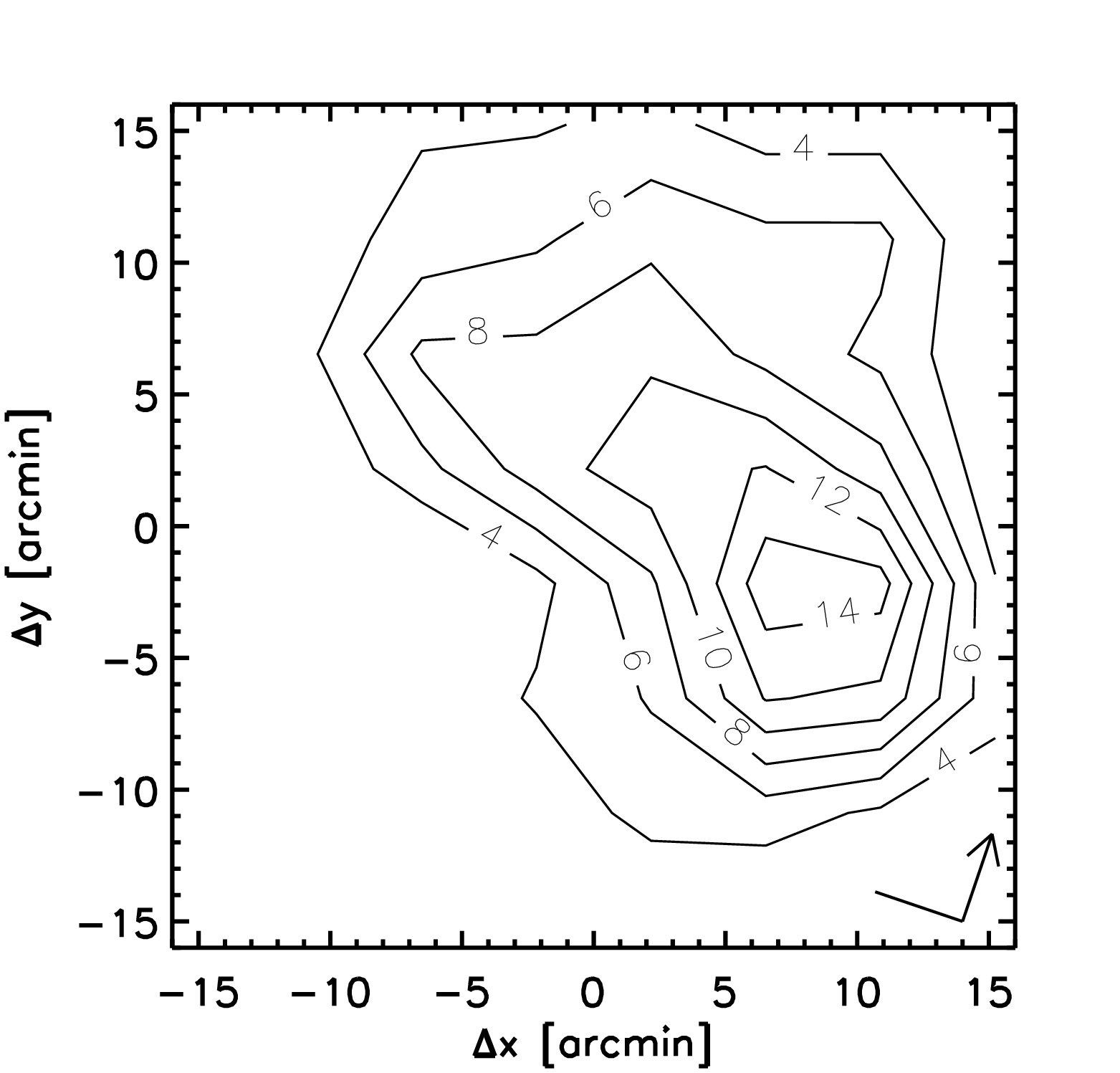,width=75mm,angle=0}
\hspace*{0.1cm}\psfig{figure=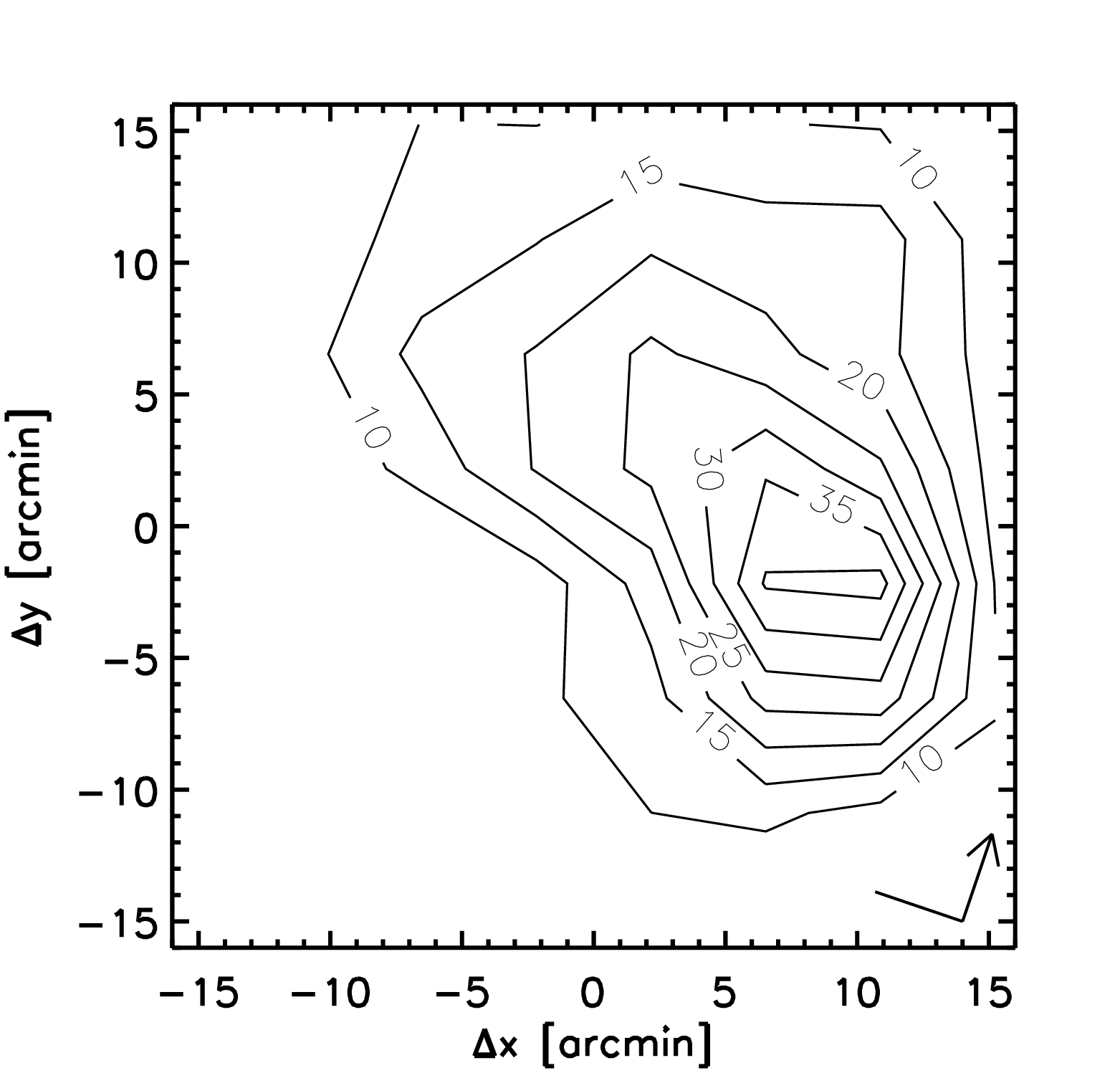,width=75mm,angle=0}}
\vspace*{0.3cm}
\caption{{\em Top-left}: Contours in MJy sr $^{-1}$ corresponding to the 60 
$\mu$m  emission from the warm component; {\em top-right}: 100 $\mu$m emission 
from the warm component; {\em bottom-left}: 100 $\mu$m  emission from the cold 
component; {\em bottom-right}: 200 $\mu$m  emission from the cold component.    
\label{fig:figure6}}
\end{figure*}

We have determined the emission maps of the warm and cold components using
the second approach described in Sect. 2.4.3. Emission maps of the warm component
at 60 and 100 $\mu$m are shown (see Fig. \ref{fig:figure6} {\em top-left} and {\em top-right},
respectively). Emission maps of the cold component at 100 and 200 $\mu$m are
shown (see Fig. \ref{fig:figure6} {\em bottom-left} and {\em bottom-right}, respectively). 
The Pearson correlation coefficients of the
different correlations are listed in Table~\ref{Table:correlations}. The
number of independent pixels (N$_{ip}$) used in the computation is also
tabulated. The PCCs corresponding to the warm and cold components individually
are generally better than the PCCs of the full emissions.

\begin{table*}
\centering
\begin{minipage}{175mm}
\caption{Pixel-pixel correlations of the far-infrared emission before
component separation (all), and for the warm and cold component. $PCC$ refers
to the Pearson correlation coefficient and $N_{ip}$ corresponds to the number 
of independent pixels. Between brackets the results for the region with 
$I^c_\nu(200)\ge$ 9 MJy sr$^{-1}$. 
\label{Table:correlations}}
\begin{tabular}{cccccccc}
\noalign{\smallskip} \hline \noalign{\smallskip}
  & \multicolumn{2}{c}{$I_\nu(25)$ vs $I_\nu(60)$} &
    \multicolumn{3}{c}{$I_\nu(60)$ vs $I_\nu(100)$} &
    \multicolumn{2}{c}{$I_\nu(100)$ vs $I_\nu(200)$} \\ 
  & all & warm & all & warm & cold & all & cold \\
\noalign{\smallskip} \hline \noalign{\smallskip}
 $N_{ip}$ & 64   & 64   &  64   & 64   & 64(31)     & 64   & 64(31)     \\
 $PCC$    & 0.88 & 0.92 &  0.96 & 0.99 & 0.93(0.88) & 0.98 & 0.98(0.97) \\
\noalign{\smallskip}\hline\noalign{\smallskip}
\end{tabular}
\end{minipage}
\end{table*}

\subsection{Colour temperature}

\subsubsection{Average colour temperatures in the illuminated and shadowed sides}

\begin{figure*}
\vspace*{-0.3cm}\centerline{
\hspace*{0.1cm}\psfig{figure=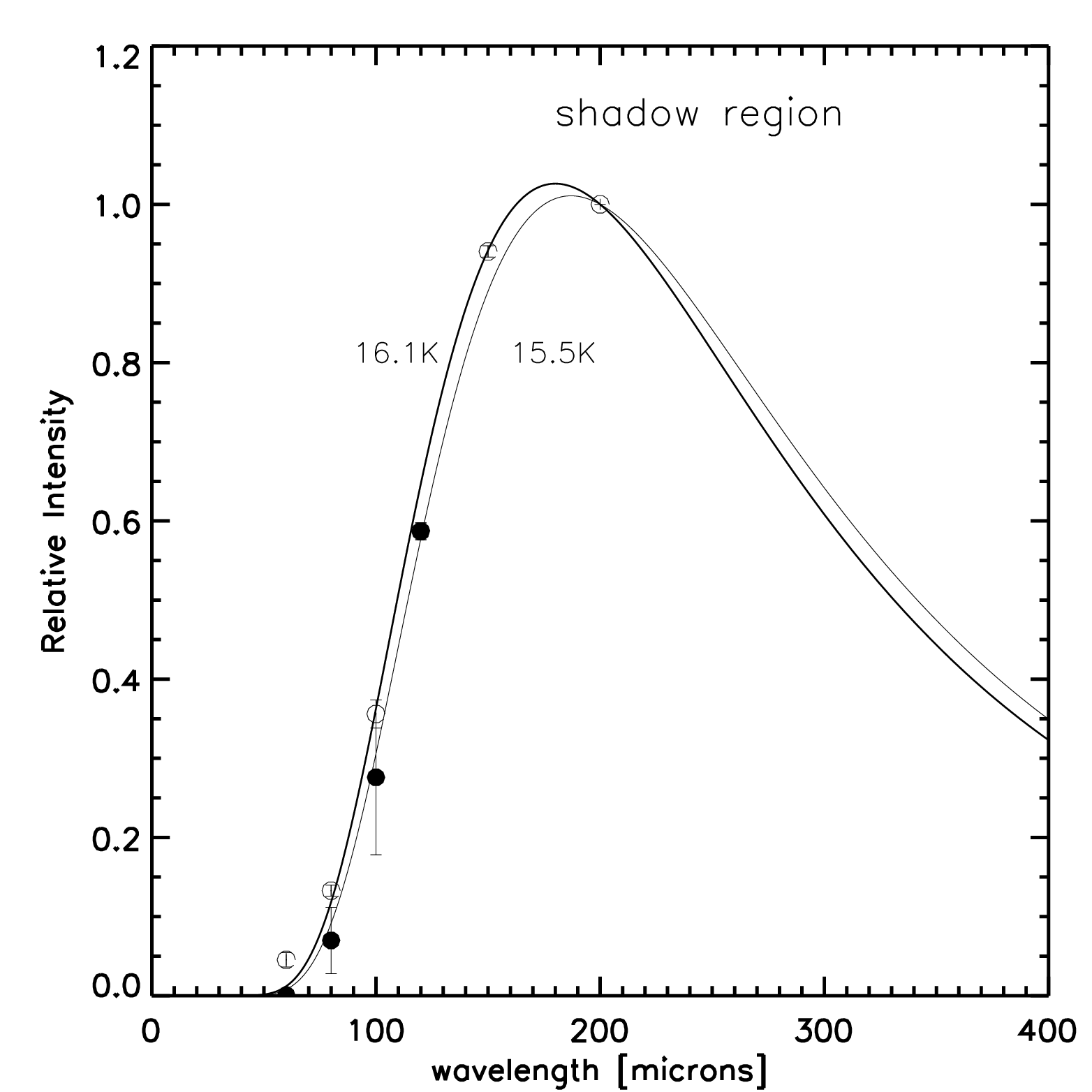,width=85mm,angle=0}
\hspace*{0.1cm}\psfig{figure=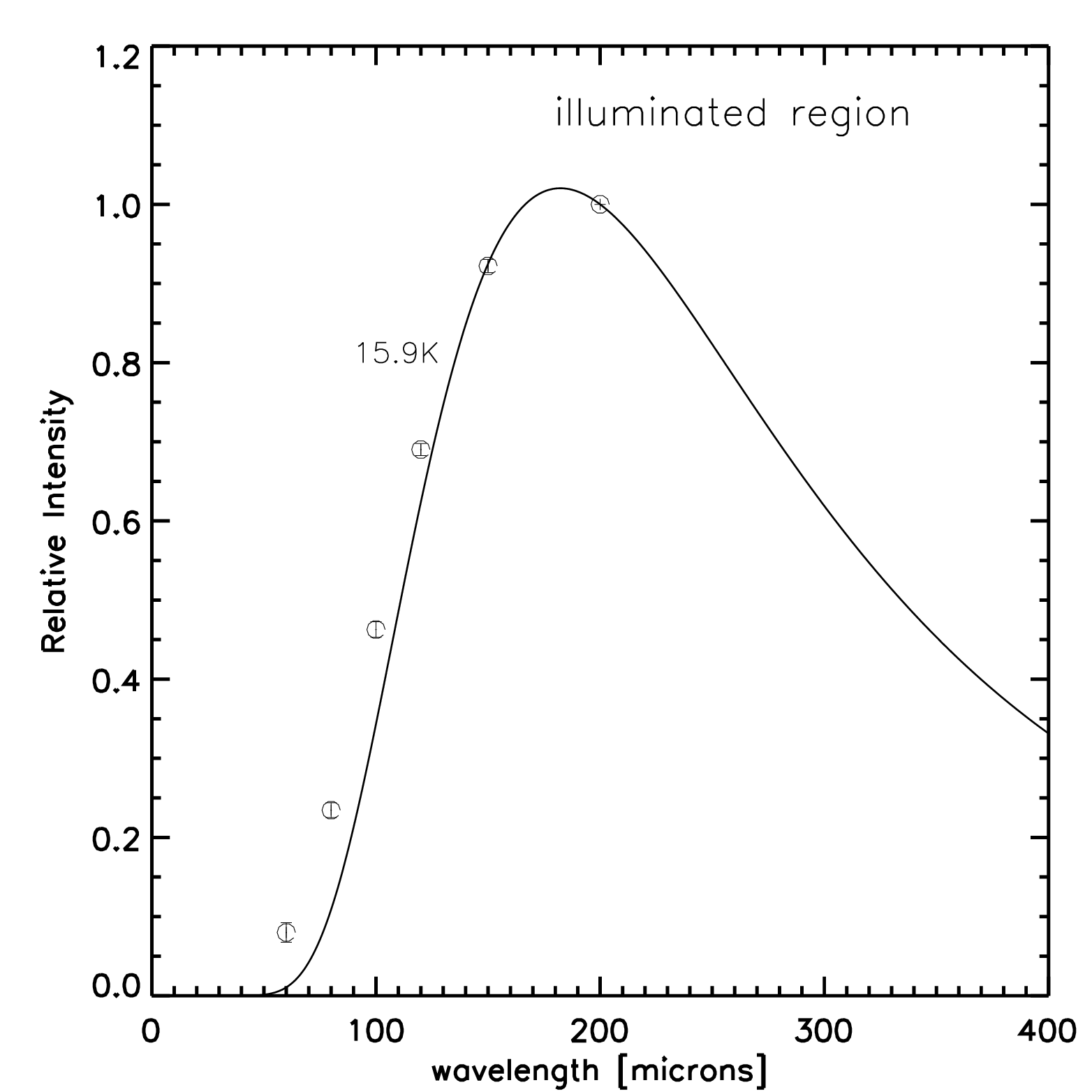,width=85mm,angle=0}}
\vspace*{0.3cm}
\caption{{\em Left}: SED in the shadowed region of LDN 1780. Two modified black bodies are overplotted: one with the colour temperature fitted to the $I_\nu(150)-I_\nu(200)$ correlation; the other one fitted to the $I_\nu(120)-I_\nu(200)$ representative of the denser region; {\em Right}: SED in the illuminated region.    
\label{fig:figure7}}
\end{figure*}

We have determined the average colour temperatures in the shadowed and illuminated sides
of LDN 1780 from the corresponding colour-corrected slope of correlation 
$I_\nu(150)-I_\nu(200)$, assuming that the spectral energies distributions are well 
represented by a modified black body with power-law emissivity $\beta=2$ as performed in Paper I. 
We have overplotted the fitted modified black body on the spectral energy distributions 
(SEDs) of the shadowed region (see Fig. \ref{fig:figure7}, {\em left}) 
and illuminated region (see Fig. \ref{fig:figure7}, {\em right}). In the shadowed region
we observed a bimodal behave, with a flattening in the correlations
$I_\nu(\lambda)-I_\nu(200)$ at higher 200 $\mu$m emission. Two modified black bodies 
with a power-law emissivity $\beta=2$ have been considered: one which fits to the 
$I_\nu(150)-I_\nu(200)$ correlation; other one that fits to the $I_\nu(120)-I_\nu(200)$ 
correlation, which is representative of the denser region 
(using $slope_2$ at 120 $\mu$m of Table \ref{isophot:slopes}). The colour 
temperatures of the fits are 16.1 and 15.5\,K, respectively. It is observed that the fitted 
models for the shadowed region reproduce within the errors the entire SED for both datasets, 
and therefore no significant excess is observed at any wavelength apart from 60 $\mu$m in the low density region. An uncertainty in the temperature of 0.2\,K has been determined from the statistical and responsivity variations.

In the illuminated region a colour temperature of 15.9$\pm$0.2\,K is obtained from the colour
ratio derive from 150 and 200 $\mu$m emissions. It is observed an emission excess at all wavelengths
including 100 and 120 $\mu$m with respect the fitted modified black body with T=15.9 and $\beta=2$.

\subsubsection{Colour temperature of the cold component}

\begin{figure}
\vspace*{0.cm}\centerline{
\hspace*{1.25cm}\psfig{figure=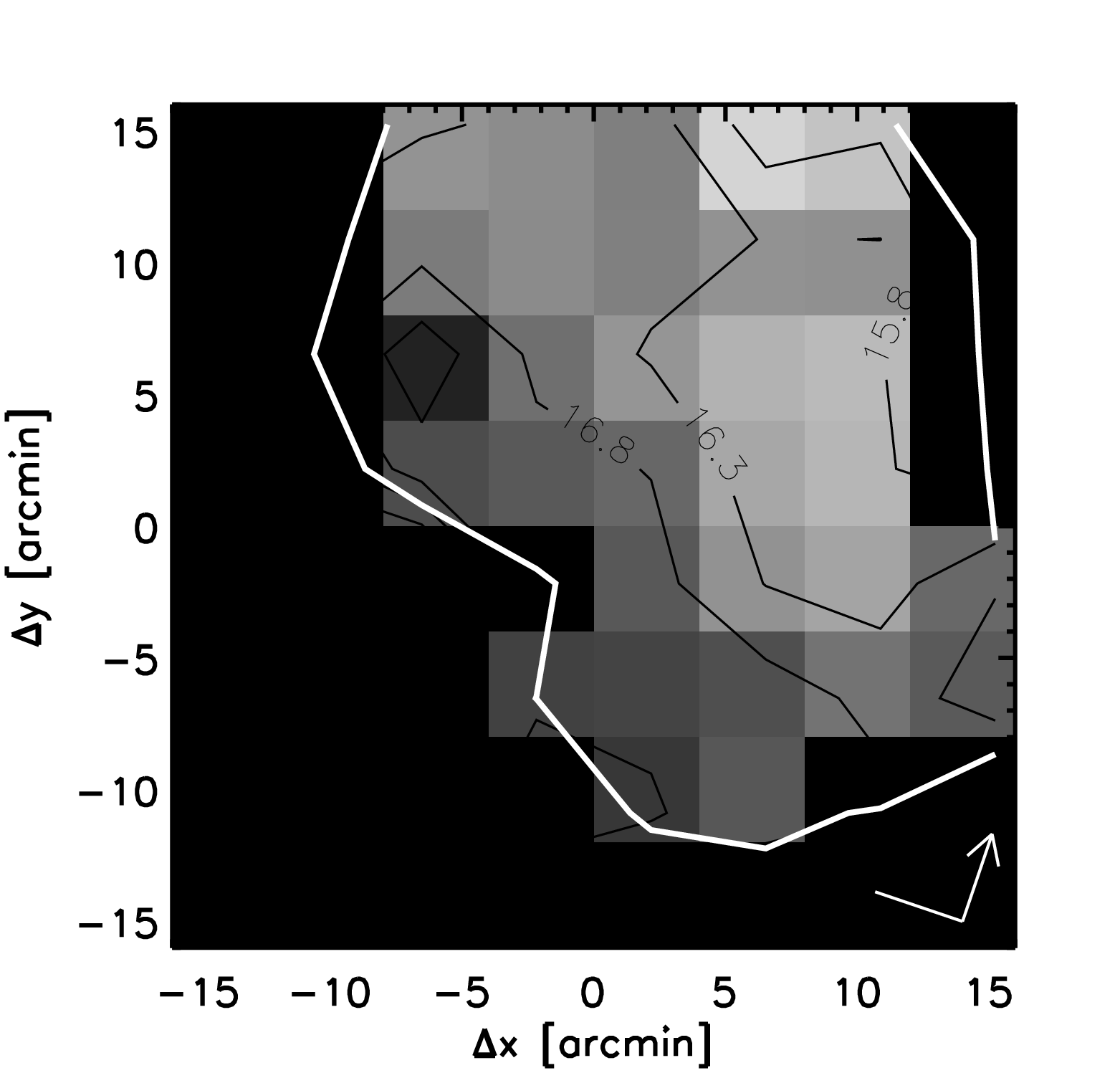,width=10cm,angle=0}}
\vspace*{0.cm}
\caption{Colour temperature map and contours (in steps of 0.5\,K) of the cold 
component. Pixel size of $4^{\prime}\times4^{\prime}$. The uncertainties $\pm$0.5\,K. 
The white thick contour marks the 9 MJy sr$^{-1}$ isophote of $I^c_{\nu}(200)$.
\label{fig:figure8}}
\end{figure}

A colour temperature map of the cold component (see Fig. \ref{fig:figure8}) was determined
from the $I^c_\nu(100)$ and $I^c_{\nu}(200)$ maps that were obtained using
the second approach described in Sect. 2.4.3. For each independent pixel, 
the emissions were colour-corrected and fitted to a modified black body with 
$\beta$ = 2. The temperature map might present a small gradient with temperatures
increasing from $\approx$15.8 to 17.3\,K along the NW-SE axis for the region
enclosed by the 9 MJy sr$^{-1}$ isophote of $I^c_{\nu}(200)$ (see Fig. \ref{fig:figure8}),
outside which the temperature errors significantly increase. The mean temperature is
16.3\,K.

The error in temperature is determined from the instrumental errors of the
{\em IRIS} 100 $\mu$m and {\em ISOPHOT} 200 $\mu$m filter-bands, as well as
the relative errors and the zero-level uncertainty (see Sect. 2.4.4). The
errors in temperature are $\pm$0.5\,K in the area closed by the isophote
$I^c_\nu(200)$ above 9 MJy sr$^{-1}$. The temperature errors at lower emissions increase to
few \,K and we have not included them in our analysis.

\subsubsection{Colour temperature of the warm component}

The colour temperature map of the warm component was determined from the
$I^w_\nu(60)$ and $I^w_\nu(100)$ maps. The
carriers in the wavelength range between 60 and 100 $\mu$m are generally VSGs 
and ``classical'' BGs. However, $I^w_\nu(60)$ was built up
to be free from VSGs and cold BGs (see Sect. 2.4.3).
For each independent pixel, emissions were colour-corrected and fitted to a modified black 
body with $\beta$ = 2. The mean colour temperature is 25\,K. 
We determined errors in temperature of $\pm$1\,K with 
the same criteria applied for the cold component but using the errors in the 
60 and 100 $\mu$m emissions. When assuming a lower value of $\beta$, temperature
increases. For example, for $\beta$=1, the colour temperature of the 
warm component would be 30$\pm$1.5\,K.

\subsection{Optical depths of the warm and cold component}

\begin{figure*}
\vspace*{0.cm}\centerline{
\hspace*{0.1cm}\psfig{figure=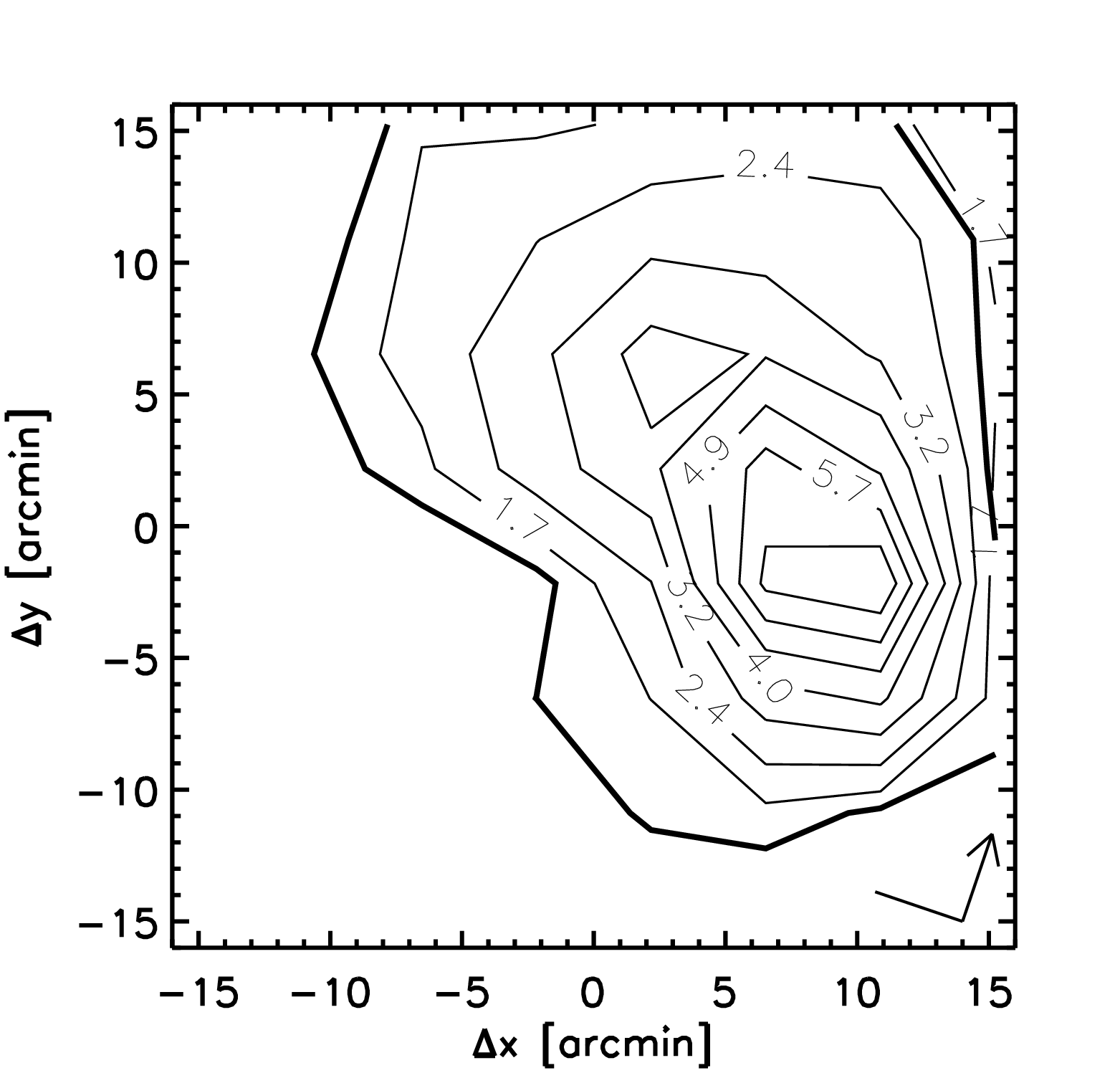,width=8cm,angle=0}
\hspace*{0.1cm}\psfig{figure=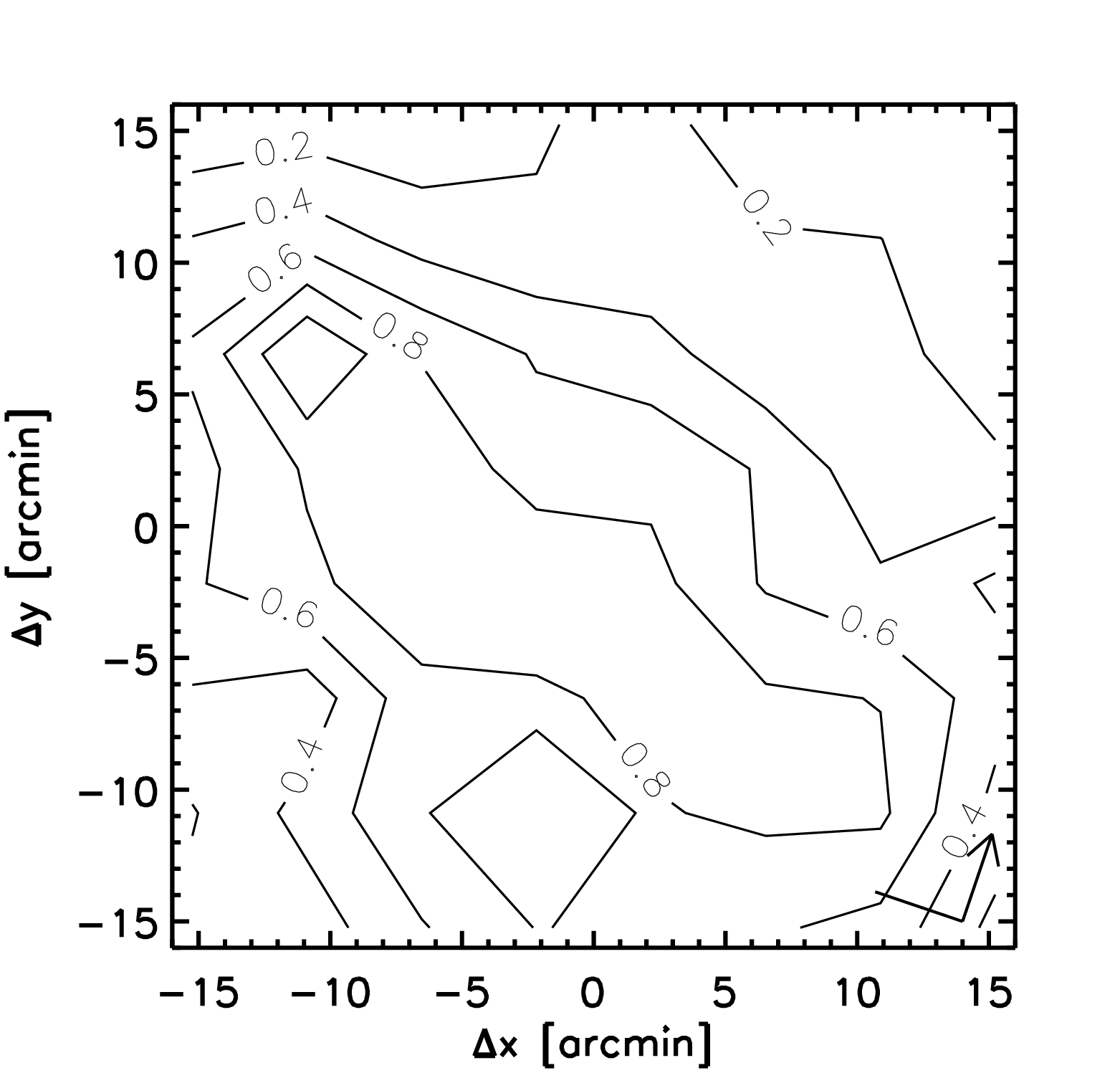,width=8cm,angle=0}}
\vspace*{0.cm}
\caption{{\em Left:} Image of $\tau^c_{\nu}(200)$ and contours overplotted in 
units of 10$^{-4}$. Contour of $I^c_\nu(200)=$ 9 MJy sr$^{-1}$ is overplotted; 
{\em Right:} Image of $\tau^w_{\nu}(100)$ and contours overplotted in units of 
10$^{-4}$. Note $\tau^w_{\nu}(100)$ can be converted to $\tau^w_{\nu}(200)$ 
dividing by 4 (see text).
\label{fig:figure9}}
\end{figure*}

The optical depth of the cold component at 200 $\mu$m (see Fig. \ref{fig:figure9},
{\em left}) has been obtained using the relation
$\tau^c_\nu(200)(x,y)$=$I^c_{\nu}(200)(x,y)/B_{\nu}(T)$, where
the Planck function $B_{\nu}(T)$ is determined at 200 $\mu$m, using the colour temperature $T$
derived in Sect. 3.2.2.
The map presents a peak at 7.3$^{+1.1}_{-0.9} \times 10^{-4}$, whose
location is in agreement with the maxima of $I^c_{\nu}(200)$ and $A_V$.
The mean errors are ${\Delta}\tau^c_\nu(200)=^{+16\%}_{-13\%}$.
 
The optical depth of the warm component at 100 $\mu$m, $\tau_\nu^c(100)$
(see Fig. \ref{fig:figure9}, {\em right}), 
has been similarly obtained from $I^c_{\nu}(100)$ and the 
Planck function at 100 $\mu$m and temperatures derived in the previous section.
We have computed an error of ${\Delta}\tau^w_\nu(100)=^{+30\%}_{-20\%}$.
The optical depth at 100 $\mu$m can be converted into optical depth at 200
$\mu$m by dividing $\tau^w_\nu(100)$ by $(200/100)^2$ assuming
$\beta=2$. We note that the optical depth of the warm component is few tens
lower than the optical depth of the cold component.

\subsection{The extinction-infrared relation: an estimation of the
emissivity of the warm component}

The correlation $I^c_\nu(200)-A_V$ (PCC=0.76) presents a higher scatter
for low emission with $I^c_\nu(200)$ below $\approx$ 10 MJy sr$^{-1}$. 
At this low emission level there are some satellite $A_V$ peaks surrounding
the $A_V$ maximum that correlate well with 
$\tau^w_\nu(100)$, i.e., the infrared tracer of the column density
of the warm component (see figures \ref{fig:figure3} and \ref{fig:figure9}, {\em right}). 
This indicates that $A_V$ also gauges the column density of the warm
component in the external regions. Therefore the total extinction can be written
as the sum the warm and cold components (see also Cambr\'esy et al. 2001):

\begin{equation}
A_V = A^c_V + A^w_V
\end{equation}

The extinction of the warm component $A^w_V$ can be related to 
$\tau^w_\nu(100)$ using the next expression:

\begin{equation}
A^w_V=\tau^w_{\nu}(100)/\frac{\tau^w_\nu(100)}{A^w_V}
\end{equation}

Assuming that the ratio $\frac{\tau^w_\nu(100)}{A^w_V}$ is constant, it is possible to 
derive $A^w_V$ from Eq. 6 and subtract it from Eq. 5 to determine $A^c_V$. 

\begin{figure*}
\vspace*{0.cm}\centerline{
\hspace*{0.1cm}\psfig{figure=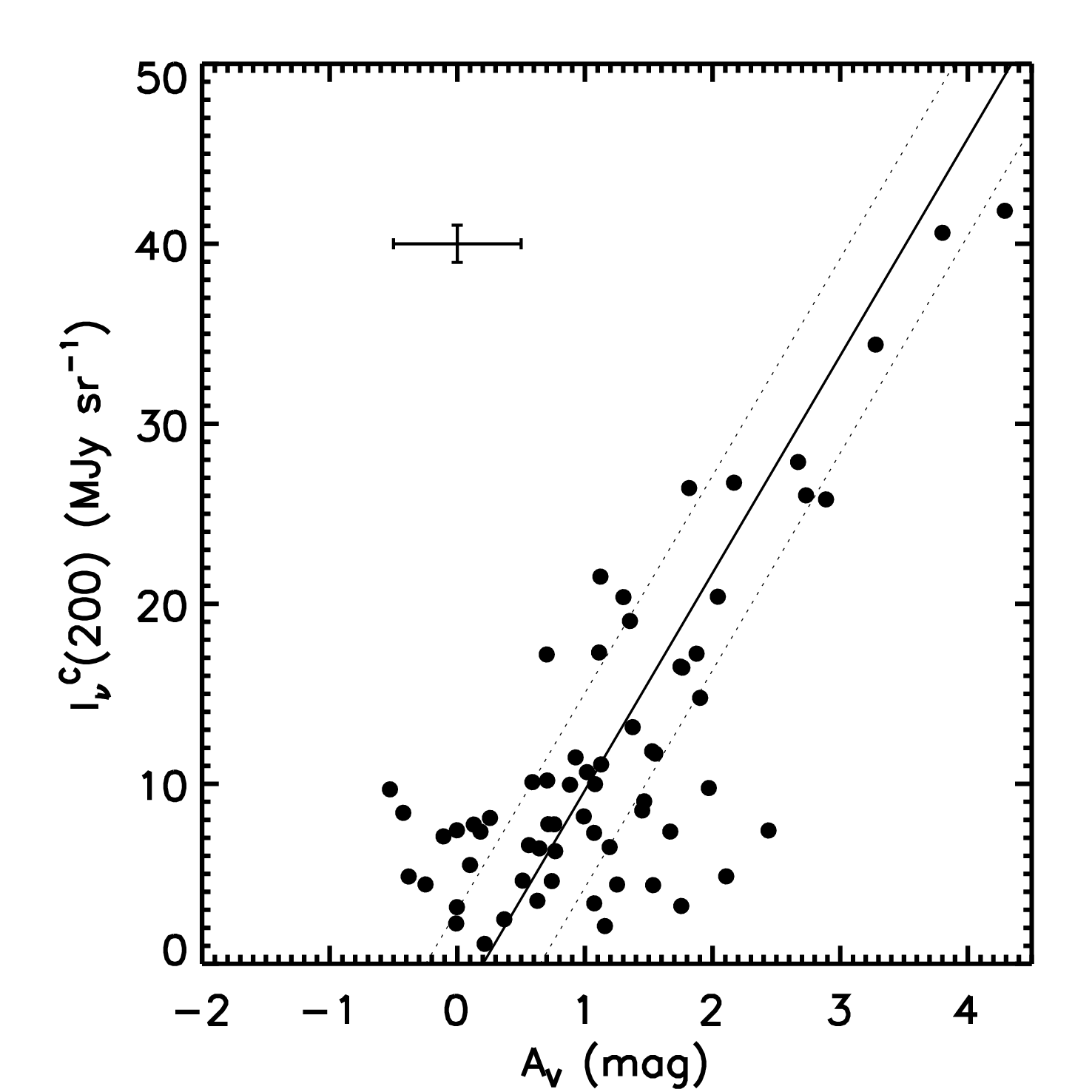,width=8cm,angle=0}
\hspace*{0.1cm}\psfig{figure=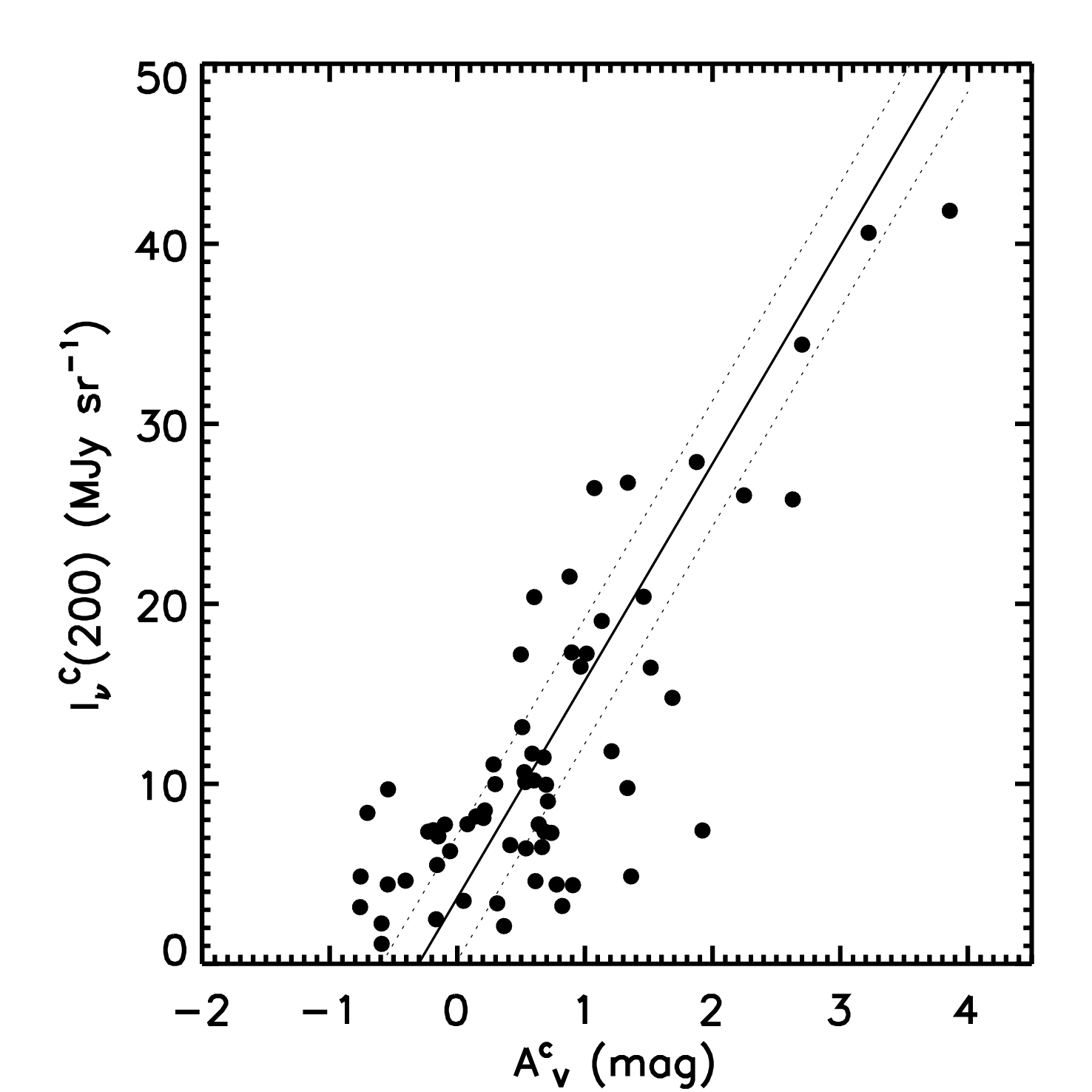,width=8cm,angle=0}}
\vspace*{0.cm}
\caption{{\em Left:} $I^c_\nu(200)$ versus $A_V$; {\em Right:} $I^c_\nu(200)$ versus $A^c_V$.
Solid lines show the corresponding least-squares fitting, and dotted lines indicate the
corresponding upper and lower 3$\sigma$ limits.
\label{fig:figure10}}
\end{figure*}

The value of $\frac{\tau^w_\nu(100)}{A^w_V}$ has been 
obtained maximising the correlation between $\tau_\nu^c$ and the residual 
$A^c_V{\equiv}A_V-A^w_V$. We obtained $\frac{\tau^w_\nu(100)}{A^w_V}\approx$1 10$^{-4}$ 
mag$^{-1}$, which corresponds to $\frac{\tau^w_\nu(200)}{A^w_V}$=0.25 10$^{-4}$ 
mag$^{-1}$ for $\beta$=2. This value implies a range of $A^w_\nu$ between 0.01 and 1.3 mag. 
For $\beta$=1, it is derived that $\frac{\tau^w_\nu(100)}{A^w_V}=0.5\ 10^{-4}$ 
mag$^{-1}$. Fig. \ref{fig:figure10} ({\em right}) shows the $I^c_\nu(200)-A^c_V$ 
correlation with PCC=0.80. It is found that PCC=0.85 if considering only
values of $I^c_\nu(200)$ above 9 MJy sr$^{-1}$.

An averaged ratio $I^c_\nu(200)/A^c_V$ = 12.1$\pm$0.7 MJy sr$^{-1}$ mag$^{-1}$ 
is derived from the $I^c_\nu(200)-A^c_V$ correlation using the bisector method 
(Isobe et al. 1990). The slope is the same when using $A_V$ instead of $A^c_V$. 
It is interesting to note that the zero level calibration established in Sect. 2.4.4
is in agreement with the origin of the $I^c_\nu(200)-A^c_V$ correlation.
Note that if adding an offset of 0.3 mag to $A^c_V$ (see Fig. \ref{fig:figure10}, {\em right}), 
the $I^c_\nu(200)-A^c_V$
will pass though the origin (i.e., $I^c_\nu(200)(A^c_V=0)$=0). 
This offset of 0.3 mag is consistent with the scatter in the relation 
$I^c_\nu(200)-A^c_V$ and also with the uncertainty in $A^c_V$ (see Sect. 2.2).

From the correlation $\tau^c_\nu(200) - A^c_V$ we have derived that
 $\tau^c_\nu(200)/A^c_V=(2.0 \pm 0.2) 10^{-4}$ mag$^{-1}$, which is similar to the value
in the DISM, but few times lower than the values found in other translucent
regions (see Paper I).

\subsection{Relations of the warm and cold components with the HI excess emission, $W_{12}$ and 
$W_{13}$}

We have observed that the warm component is located in the illuminated side of LDN 1780 as well as the HI excess emission (Mattila \& Sandell 1979).

\begin{figure}
\vspace*{0.cm}\centerline{
\hspace*{0.1cm}\psfig{figure=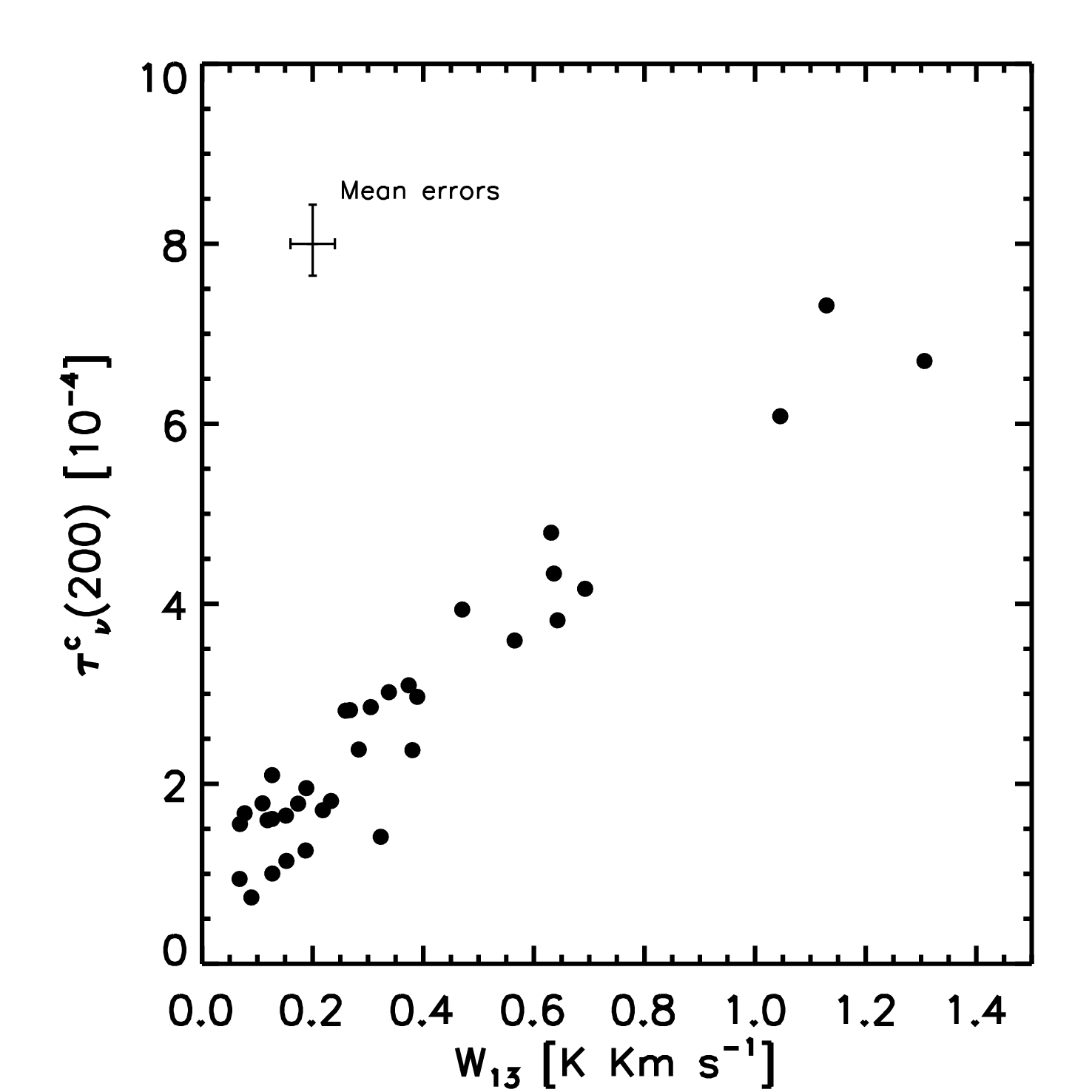,width=8cm,angle=0}}
\vspace*{0.cm}
\caption{Correlation $W_{13}-\tau^c_{\nu}(200)$. Mean 
errors are indicated. A total of 34 independent pixels with size $6^{\prime}\times6^{\prime}$ is shown.
\label{fig:figure11}}
\end{figure}

The monoxide line integrated $W_{12}$ and $W_{13}$ do not correlate with
the warm component. The correlations with the cold component are very good
(see Fig. \ref{fig:figure11} and Table~\ref{Table:correlations_13CO}). The correlation 
between $W_{13}$ and $\tau^c_{\nu}(200)$ and also with $I^c_\nu(200)$ are especially 
good with PCC=0.96. The morphologies of $\tau^c_{\nu}(200)$ and $W_{13}$ are very 
similar (see Figs. \ref{fig:figure5} and \ref{fig:figure9}, {\em left}).

The ordinary least squares bisector method has been used to determine 
$\tau^c_\nu(200)/W_{13}$ from the correlation $\tau^c_\nu(200)-W_{13}$.
We obtained that $\tau^c_\nu(200)/W_{13}=(5.2 \pm 0.2){\times}10^{-4}$ K$^{-1}$
Km$^{-1}$ s. An offset of ${\Delta}\tau^c_\nu(200)=(0.91 \pm 0.07){\times} 10^{-4}$
is obtained.

\begin{table*}
\centering
\begin{minipage}{175mm}
\caption{Correlations of $W_{13}$ and $W_{12}$ with $I_\nu(200)$, 
$I^c_\nu(200)$ and $\tau^c_\nu(200)$. $PCC$ refers to the Pearson correlation 
coefficient and $N_{ip}$ corresponds to the number of independent pixels. 
$W_{13}$ and $W_{12}$ are positive values. Between brackets the results for the 
region with $I^c_\nu(200)\ge$ 9 MJy sr$^{-1}$. 
\label{Table:correlations_13CO}}
\begin{tabular}{ccccccc}
\noalign{\smallskip} \hline \noalign{\smallskip}
   &  \multicolumn{3}{c}{$W_{13}$} &
    \multicolumn{3}{c}{$W_{12}$} \\
  & $I_\nu(200)$ & $I^c_\nu(200)$ & $\tau^c_\nu(200)$ & $I_\nu(200)$ & 
$I^c_\nu(200)$ & $\tau^c_\nu(200)$ \\
\noalign{\smallskip} \hline \noalign{\smallskip}
 $N_{ip}$ & 34   & 34(26)      & 34(26)     &  52    &  52(31)  &  51(31) \\
 $PCC$    & 0.79 & 0.96(0.95)  & 0.96(0.96) &  0.77  &  0.82(0.69)   &  
0.74(0.63)  \\
\noalign{\smallskip}\hline\noalign{\smallskip}
\end{tabular}
\end{minipage}
\end{table*}

\subsection{Discussion}

\subsubsection{Separation of the warm and cold components}

LFHMIC95 already pointed out that the profile (along the direction of the Upper 
Scorpius OB association) of 25 $\mu$m emission in LDN 1780 shows a different
distribution with respect to the 12 and 60 $\mu$m emissions. In the classical picture, 
this could be interpreted as due to a stronger emissivity of the VSGs, 
which mostly emit at 60 $\mu$m. We have alternatively interpreted that the 
spectra energy distribution in LDN 1780 is mostly due to the existence of cold BGs
together with warm BGs and VSGs illuminated by an enhanced local interstellar 
radiation field. The warm and cold components have high temperatures 
compared to other moderate density regions 
(see Paper I) and consequently the bulk of their emissions is displayed towards shorter 
wavelengths. In the illuminated side of the stripe, it is observed emission excess at 
100 and 120 $\mu$m over a modified black body with the colour temperature
derived from the $I_\nu(150)-I_\nu(200)$ correlation. This excess at such
long wavelengths can not be due to VSGs and supports the existence of an additional 
BG component.

In previous studies, the 60 $\mu$m emission has been 
used as template to derive the excess emission at 100 $\mu$m 
(Laureijs, Clark \& Prusti, 1991; Abergel et al. 1994) and also 200 $\mu$m (see Paper II).
The correlation ${\Delta}I_\nu(100)-I_\nu(200)$ (PCC=0.70) derived from our
first approach is not as good as the $I^c_\nu(100)-I_\nu(200)$
correlation (PCC=0.98) obtained from the second approach (see Sect. 2.4.3). 
Different mean colour temperatures are derived, with $\sim$13\,K and $\sim$16\,K
for the first and second approaches, respectively.
The low temperature of $\sim$13\,K is likely due to the fact that $I_\nu(60)$ contains
not only emission from VSGs but also from warm BGs. The second approach to separate
the warm and cold components is intended to sort out this problem.
The S/N of the emission at 25 $\mu$m is high enough to be used as template 
of the VSG component.

The goodness of the second approach to disentangle
the different dust components is supported by the good agreement between the temperatures obtained for the squared field map of the cold component and the shadowed side of the stripe-like region. It is also supported by the very good correlation between
W$_{13}$ and $\tau^c_{\nu}(200)$. When using the first approach the optical depth 
of the cold component is not so well correlated with W$_{13}$. The 
${\Delta}I_\nu(100)-I_\nu(200)$ correlation is not as good as $I^c_\nu(100)-I_\nu(200)$.
In addition, the warm component presents a distribution which is similar to the HI excess emission, facing the illuminated side of the cloud

Regarding the properties of the warm component (see Sects. 3.2.3 and 3.3)
it is difficult to derive its emissivity. We have found a ratio $\tau\nu^w(100)/A_V$
which is significantly low. The temperature could be also somehow overestimated since
it could be that $I_\nu(200)$ contains some contribution of the warm component, which
is not considered in Eq. (4).
However, the mentioned agreement between the temperature derived for the stripe in the
shadowed region and temperature in the corresponding region in the squared field supports
the consistency of our results.

\subsubsection{Properties of the cold component}

The cold component exhibits a temperature of $\approx$16\,K, which is higher 
than expected at the UV-shielded densities of 10$^2$-10$^3$ cm$^{-3}$. This 
requires either an extra heating rate through the cloud or a change in the 
properties of the grains with respect those of the DISM.

The average value of $I_{\nu}(200)$/A$_V$=12.1$\pm$0.7 MJy sr$^{-1}$ mag$^{-1}$ 
of the cold component in LDN 1780 is below the value of the diffuse 
interstellar medium with $I_{\nu}(200)$/A$_V$=21.1 MJy sr$^{-1}$ mag$^{-1}$ and 
T=17.5\,K, and also below the values of the moderate density regions of the 
sample in Paper I. These regions are generally colder than LDN 1780 and
have a cold component with an enhanced far-infrared emissivity.

The average ratio $\tau^c_{\nu}(200)/A^c_V$=(2.0$\pm$0.2)$\times$10$^{-4}$ mag$^{-1}$ is slightly lower 
than the value of the DISM ($\tau^c_{\nu}(200)/A_V$=2.4$\times$10$^{-4}$ mag$^{-1}$) and the
same than in the warmer (T=18.9$\pm$0.7\,K) G90.7+38.0 (2.0$\pm$0.8)$\times$10$^{-4}$ mag$^{-1}$)
(see Paper I). When considering values of independent pixels, we find a variation
in $\tau^c_{\nu}(200)/A^c_V$ from 1\ 10$^{-4}$ mag$^{-1}$ to 3\ 10 $^{-4}$ mag$^{-1}$ for temperatures of
17.3 and 15.8 \,K, respectively. A small enhancement in the emissivity of the
cold component is therefore observed. This could be related to grain coagulation
as observed in other translucent clouds (see Paper I).

\subsubsection{Properties of the warm component}

The temperature of the warm component of 25$\pm$1\,K ($\beta=2$) is higher than 
the average value of the DISM
(17.5\,K). The warm component, which is surrounding the cold component, it is certainly
affected by the anisotropy of the radiation field. The maximum of the far infrared emission
of the warm component is located in the illumination side of the cloud.

For $\beta$=2 we obtained $\tau^w_{\nu}(200)/A^w_V$=0.25$\times$10$^{-4}$ mag$^{-1}$. This value
is significantly lower than the values for the cold component as well as the DISM and
G90.7+38.0. All these regions are significantly colder than the warm component in LDN 1780.
This supports that the properties of the dust grains of the warm component are different
from those of the DISM as well as the cold component. This could be result of the
enhanced radiation field in the surroundings of LDN 1780 with respect the DISM. The
observation of the emission of all the area LDN 1780 at intermediate wavelengths 
between 100 and 200$\mu$m would help to a better extraction of the warm component.

Note that $\beta$ could be different for the warm and cold components (see also
discussion in Paper II). An unrealistic value of $\beta$=1 for the warm component
implies that $\tau^w_{\nu}(100)/A^w_V$ decreases $\sim$50\%. If $\beta\sim$1.5, then
T$\sim$27\,K and the ratio $\tau^w_{\nu}(100)/A^w_V$ drops $\sim$20\% with respect
to the value for $\beta$=2. A further investigation is out of the scope of the data
presented here.

\section{Ionisation process}
\subsection{Relation $A_V-I_{\nu}({\rmn H}\alpha)$}

\begin{figure*}
\vspace*{0.cm}\centerline{ \hspace*{0.1cm}
\psfig{figure=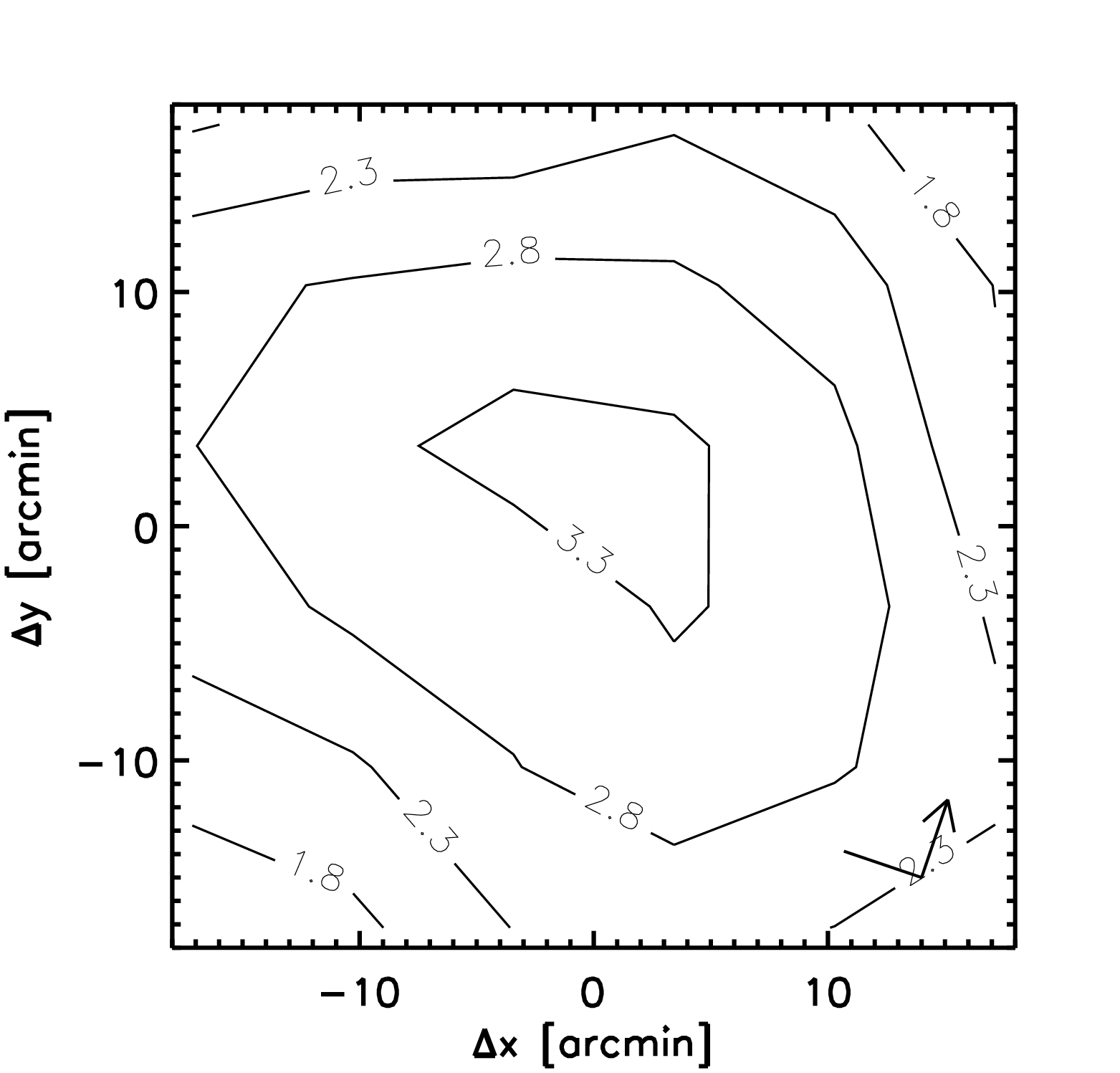,width=8cm,angle=0} 
\hspace*{0.1cm}\psfig{figure=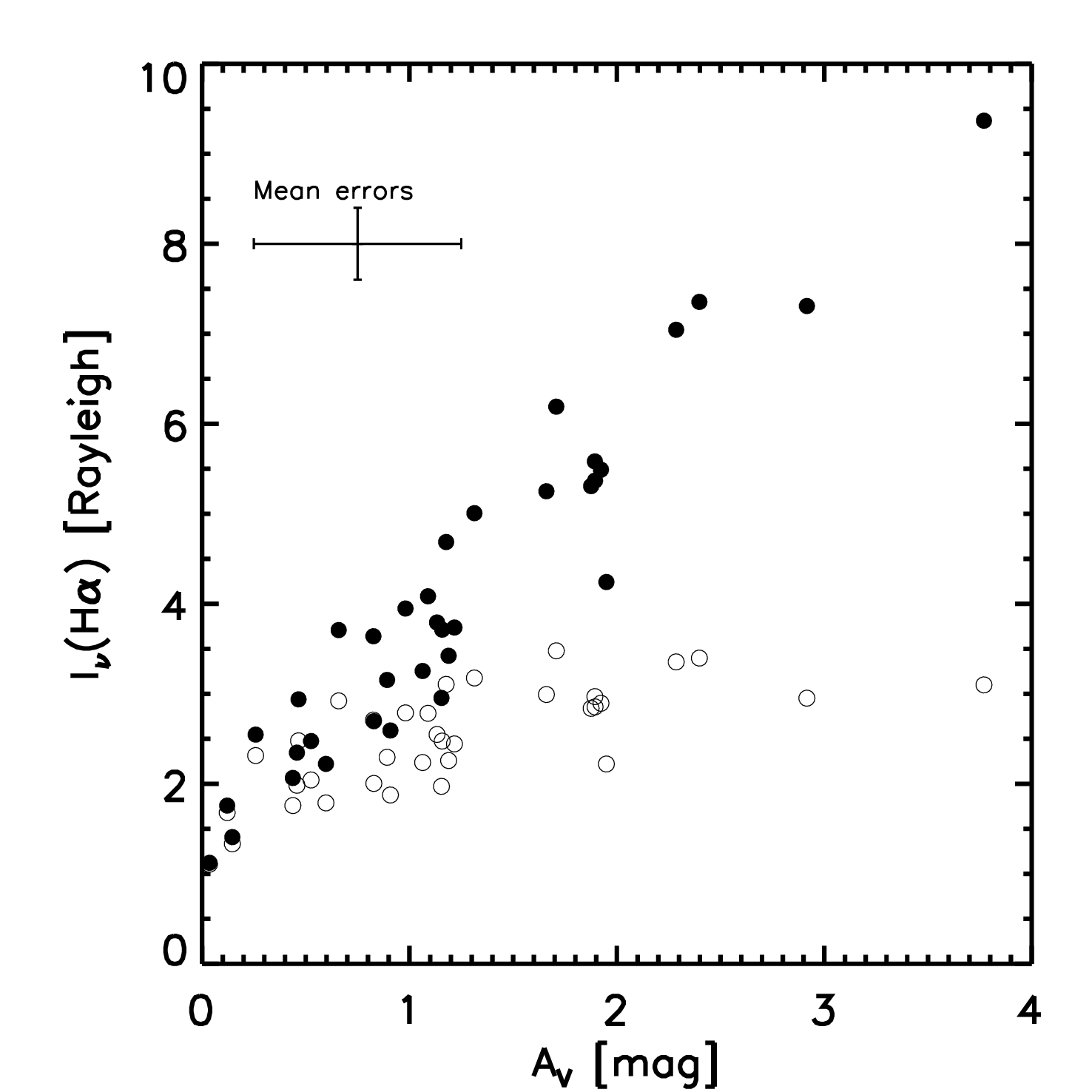,width=8cm,angle=0}}
\vspace*{0.cm}
\caption{{\em Left:} $I_{\nu}({\rmn H}\alpha)$ contours (before correction
from extinction) in Rayleigh; {\em Right:} $I_{\nu}({\rmn H}\alpha)$ before (open circles)
and after (filled circles) correction from extinction versus $A_V$. A total
of 36 independent pixels with size of $6^{\prime}\times6^{\prime}$. Mean
error bars are also indicated.
\label{fig:figure12}}
\end{figure*}

The H$\alpha$ emission in LDN 1780 (see Fig. \ref{fig:figure12}, {\em left}) varies 
between $\sim 1$ and 4 Rayleigh, and its error is $\sim$0.4 Rayleigh (Finkbeiner 2003).
The extinction $A_V$ correlates very well with the H$\alpha$ emission, with a Pearson
correlation coefficient of 0.71 
(see Fig. \ref{fig:figure12}, {\em right}). There is a linear relation between the H$\alpha$ 
emission and $A_V$ that remains till $A_V\sim$2 mag, and then flattens. The correlation is
observed in regions with an abundant molecular gas content, which is surprising since the 
UV radiation field from the Galactic plane and the Sco-Cen OB association cannot come in through so deep (see Sect. 4.2.1).

Such a good correlation between the H$\alpha$ emission and the extinction 
in LDN 1780 strongly suggests that the ionised gas and the dust 
share a significant emitting volume and are quite uniformly mixed within it.
The flattening in the H$\alpha$ emission at high $A_V$ ($\sim$2 mag) is 
likely result of dust extinction (see Fig. \ref{fig:figure12}, {\em right}).
It is therefore suitable to correct the observed H$\alpha$ emission
$I_{\nu,obs}({\rmn H}\alpha$) from extinction using the following equation:

\begin{equation}
I_{\nu}({\rmn H}\alpha)=\frac{I_{\nu,obs}({\rmn H}{\alpha})~\tau({\rmn H}{\alpha})}{(1-e^{-\tau({\rmn H}{\alpha})})}
\end{equation}

The optical depth at the wavelength of H$\alpha$ is obtained from our
near-infrared extinction map and the Cardelli, Clayton \& Mathis (1989)
extinction law which gives $A_{{\rmn H}\alpha}/A_V = 0.72$ so that
$\tau({\rmn H}\alpha) = 2.5 \log e \times 0.72 \times A_V$.
In the surroundings of the cloud, with $\tau({\rmn H}{\alpha}) \rightarrow 0$,
the equation (6) gives
$I_{\nu}({\rmn H}\alpha) \approx I_{\nu,obs}({\rmn H}\alpha)$; 
toward the peak of the cloud 
$I_{\nu}({\rmn H}\alpha) \approx \tau({\rmn H}{\alpha})\ I_{\nu,obs}({\rmn H}\alpha)$.

The $A_V-I_{\nu}({\rmn H}\alpha)$ plot, with the extinction-corrected
H$\alpha$ is shown in Fig. \ref{fig:figure12} ({\em right}). Using a ordinary least squares
bisector method (Isobe et al. 1990), we obtained a ratio
$I_{\nu}({\rmn H}\alpha)/A_V$=2.2$\pm$0.1 R mag$^{-1}$.
This is twice the value of $I^{obs}_{\nu}({\rmn H}\alpha)/A_V$.
Note there is an offset in the relation with 
$I_{\nu}({\rmn H}\alpha)$=1.4$\pm$0.15 R at $A_V=0$ mag. 

\subsection{Enhanced ionisation rate}

The disk-like distribution of emitting gas in
the Galaxy produces an emission $I_{\nu}({\rmn H}\alpha) \approx \frac{1}{sin(b)}$ Rayleigh
(Reynolds 1984), which corresponds to 1.7 R at the latitude of LDN 1780.

The observed relation between $I_{\nu}({\rmn H}\alpha)$ and $A_V$ indicates that
the ionisation and recombination processes are even produce in very deep regions
of the cloud, with optical extinctions of $\sim$2 mag. This is likely due to
a source of ionisation that can penetrate much deeper into the cloud than the UV and 
the X-ray radiations and produces a nearly uniform ionisation rate in the cloud.

The observed ratio $I_{\nu}({\rmn H}{\alpha})/A_V$ gives us the possibility 
to estimate the ionisation rate from the following equation:

\begin{equation}
\xi\approx\frac{\frac{I_\nu(H_\alpha)}{A_V}}{\frac{N(H+H_2)}{A_V}}
\end{equation}

which results to be $\sim$10$^{-16}$  $\gamma$ s$^{-1}$. We have used
$N(H+H_2)/A_V$ = $1.87 \times 10^{21}$ mag$^{-1}$ cm$^{-2}$ (Bohlin, Savage \& Drake 1978).
Although this corresponds to a very high ionisation rate for the
intermediate density cloud LDN 1780, it worth noting it is only a lower limit
obtained by assuming that every collision with an high energy particle yield
a H$\alpha$ photon emission.
Other clouds in the LDN 134 like LDN 169, LDN 183 and LDN 134 also present
H$\alpha$ emission. The origin of the ionisation source is discussed in the
next sections.

\subsubsection{Influence of the X-ray and UV radiation fields}

LDN 1780 as well as LDN 134, LDN 183 and LDN 169 are likely influenced 
by an anisotropic radiation field (THLM95, LFHMIC95). The southern sides of 
these clouds face the nearby UV sources $\zeta$ Oph (an O9.5V star at a 
distance of 140 pc from the Sun) and the Upper Scorpius OB association.

LFHMIC95 studied the variation of the dust and molecular gas tracers along a 
line parallel to the direction of the UV radiation field (assumed toward Upper 
Scorpius OB association) crossing the centres of LDN 1780 and LDN 134. They 
pointed out some differences between these clouds. For LDN 1780: i) no 
detection of dense and opaque clumps is observed, although THLM95 
found a weak emission of C$^{18}$O, ii) the emission is significantly brighter 
at 25, 60 and 100 $\mu$m emission compared to LDN 134 (also LDN 183); iii) 
the 60 and 100 $\mu$m profiles are asymmetric (as in LDN 134), but similar, 
and peak at the same position as the carbon monoxide lines, while in LDN 134 
there is an offset for the W($^{13}$CO); iv) $I_{\nu}(60)/I_{\nu}(100)$ and 
$I_{\nu}(12)/I_{\nu}(25)$ decreases toward the shadow side, while in LDN 134 
they are either nearly constant or present a decrease toward higher
densities; v) the CO temperature peak is highly symmetric, contrary to
LDN 134; vi) the lines are narrower on the shadow side.

THLM95 suggested that only the external region of LDN 1780 might be affected by 
the local radiation field. They estimated that the radiation field
in the illuminated side of LDN 1780 could few (2 or 3) times higher than 
the average radiation field at its Galactic latitude because of the presence of 
the OB association. THLM95 concluded the asymmetric UV illumination from the 
Scorpius-Centaurus association and from the Galactic plane has influenced the 
morphology of LDN 1780, resulting in the offset of the HI emission maximum, the 
blue-shifted (out-flowing) molecular gas envelope and the enhancement of the 
VSG abundance toward the side facing the Galactic plane. THLM95 
claim that the offset found between the radiation temperature peaks of CO and $^{13}$CO 
and the comet-like morphology cannot be explained by the UV asymmetry. They concluded
from the study of the morphologies of CO(J=1-0) and $^{13}$CO(J=1-0) that 
the offset between the maximum of the CO temperature peak and the geometrical 
centre can be explained as result of a shock front that passed through the 
cloud. The core of $^{13}$CO and the evaporating lower density envelope of gas 
pointing to NE correspond to the ``head'' and the 
``tail'' of the cometary nebula, respectively (see next Sect.).

The maps of H$\alpha$ emission and $A_V$ show a good correlation, and no 
asymmetry evidence is observed between the shadow and illuminated sides. 
The observe H$\alpha$ enhancement is entirely related to LDN1780. We also 
observe that H$\alpha$ emission is present in the other clouds of the LDN 134 complex.
However, LDN 1780 is the only cloud of the LDN 134 complex where ERE emission
has been reported (Chlewicki \& Laureijs 1987). Smith \& Witt (2002) have
interpreted the ERE in LDN 1780 as result of photoluminescence of silicon nanoparticles.
They observed a relatively high quantum efficiency of the ERE process ($\sim$13\%) 
as well as a long wavelength emission peak at 7000 \AA. They argued that
given the relatively high electron density and low UV photon density within 
the cloud in comparison with the DISM, the overall 
ionisation equilibrium tends more towards neutrality, and then the silicon
nanoparticles are predominantly neutral, with an increment in the abundance
of the largest particles, and then the overall quantum efficiency increases
and the emission peak shifts to longer wavelengths. We have checked the H$\alpha$
emission maps and far-infrared emission of the 23 objects (DISM, LDN1780, reflection nebulae,
HII regions) in the sample of Smith \& Witt (2002) finding that there are only
few of them with a clear counterpart in H$\alpha$. LDN 1780 is one of the few of them
with a correlation between H$\alpha$ and far-infrared emission. We have also
noticed that other clouds of the LDN 134 complex as the dense LDN 183 present H$\alpha$
in regions with a very low emission at 12 and 25 $\mu$m.

We conclude that the asymmetry of the radiation field is mainly affecting 
the warm component, which mostly contain {\em warm} BGs. 
The warm component is likely associated to the HI excess emission
and it is surrounding the cold component (see Figure \ref{fig:figure6}). 
The cold component, consisting of cold big dust grains, is well shielded from the UV radiation.
The observation of H$\alpha$ emission at high column densities where CO emission 
is also observed supports the existence of a ionisation flux that can 
penetrate very deep into the cloud reaching the inner regions of it with $A_V\approx$2 mag. 
The UV radiation field from the Galactic plane and the OB association significantly
decreases for $A_V\approx$0.3-0.5 mag and can not explain the observed CO and
H$\alpha$ emissions. 
The X-rays dominates the cosmic ionisation for smaller column densities. 
Therefore, both ionisation sources cannot give account of the observed ionisation rate
in LDN 1780 and can be rejected out.

\subsubsection{Is it possible that the H$\alpha$ emission in LDN 1780 has been 
produced by a shock?}

THLM95 pointed out the importance of a shock front of $\sim$10 Km s$^{-1}$ in 
the morphology of LDN 1780. Based on the idea of de Geus et al. (1989), THLM95 
suggested that an expanding HI shell originated from the Scorpius-Centaurus OB 
association could be eventually fragmented forming clumps that were reached by 
a shock front (of 10-15 km s$^{-1}$) of supernovae originated in the Upper 
Centaurus-Lupus subgroup of Sco-Cen association. As result, some cometary 
shaped clouds as LDN 1780 were formed.

Assuming that LDN 1780 was formed as result of a shock front, is it 
possible that the emission in H$\alpha$ has been {\em  induced by a shock}? There
is an evolution in the H$\alpha$ surface brightness,
$I_{\nu}({\rmn H}{\alpha})$, of a region that undergoes a shock
and compress to form a molecular cloud.
During the initial phase, when the gas temperature is still high
($\sim$ 1000\,K) the typical surface brightness of H$\alpha$ is
$\approx 0.005$ (0.009) R for a shock speed of 10 (20) Km s$^{-1}$.
In general, once H$_2$ is formed (dominated by self-shielding or dust
shielding depending on the shock speed, see Bergin et al. 2005) nearly
all hydrogen is molecular (only about 1\% by fraction in atomic form).
Only a small fraction of H$^+$ is produced, which will give H$\alpha$ when 
it recombines. This should drastically reduce the H$\alpha$ emission once the 
gas is (mostly) molecular (Edwin Bergin 2005, private communication). Higher 
values of the shock speed front yield higher values of
$I_{\nu}({\rmn H}{\alpha})$ at the first stages of the shock,
but $I_{\nu}({\rmn H}{\alpha})$ always drops by several orders once
H$_2$ is formed. The timescale of CO molecular cloud formation is not
established by the H$_2$ formation rate on to grains, but by the timescale
for accumulating a sufficient column density ($A_V \ge 0.7$ mag;
Bergin et al. 2005).

\subsubsection{A plausible solution: confinement by self generated MHD waves of 
cosmic rays}

We interpret the observed ionisation rate for LDN 1780, 
$\xi \sim 10^{-16} \gamma$ s$^{-1}$, as due to an enhanced cosmic ray
flux. The observed value of $\xi$ is around one order of magnitude higher
than the standard value for the ISM ($\xi \sim 10^{-17}$ s). 
Our determination of $\xi$ depends on the assumed standard ratio $A_V/N_{\rmn H_2}$. 
This ratio could depend on the environment conditions; 
a factor 2 for the uncertainty is probably conservative.

Enhanced values of $\xi$ have been suggested for the DISM by McCall et al. (2003) from
H$_3^+$ measurements, but it is by first time found in a translucent cloud.

Low-energy cosmic rays of $\sim$100 MeV dominate the ionisation fraction and 
heating of cool neutral gas in the interstellar medium, especially in dark 
UV-shielded molecular regions (Goldsmith \& Langer 1978). The observed 
H$\alpha$ enhancement could be explained as result of a confinement by self 
generated MHD waves of low energy cosmic rays (Padoan \& Scalo 2005). There are 
other alternatives that require further investigation as multiple magnetic 
mirrors that could lead to cosmic ray variations (Cesarsky \& Volk 1978).

The study of Padoan \& Scalo (2005) indicates the possible enhancement of the 
cosmic ray flux for densities below $\sim$5$\times$10$^2$ cm$^{-3}$. Although 
this value depends on different assumptions (see Padoan \& Scalo 2005 for 
discussion), it is in good agreement with the value found for the $^{13}$CO core of 
LDN 1780, with a density of $6\times 10^{2}$ cm$^{-3}$ according to THLM95, and
$10^3$ cm$^{-3}$ for LFHMIC95. Our relatively high temperature 
($\sim$16\,K) for the cold component, whose emission correlates very well with 
$A_V$ and $^{13}$CO, also supports the presence of a source of ionisation 
like the cosmic rays that go much deeper than UV photons and contribute to
the heating of the cloud.  

The interaction of cosmic rays with molecular clouds can produce H$\alpha$
by several ways. The more direct one would be
$CR + {\rm H}_2 \rightarrow {\rm H} + {\rm H}^+ + e^-$
where the H$^+$ emits H$\alpha$ by recombination with an electron.
The result of the reaction also contains a neutral H which can be
excited and lead to an H$\alpha$ photon emission. The collision of
an energetic particle with molecular hydrogen does not necessarily
produce an hydrogen atom. 
The reaction could be $CR + {\rm H}_2 \rightarrow {\rm H}_2^+ + e^-$
which reacts quickly with another H$_2$ (Lepp 1992):
${\rm H}_2^+ + {\rm H}_2 \rightarrow {\rm H}_3^+ + {\rm H}$
followed by the classical dissociative recombination
${\rm H}_3^+ + e^- \rightarrow {\rm H} + {\rm H} + {\rm H}$
or ${\rm H}_2 + {\rm H}$.
All the hydrogen atoms resulting from these reactions can potentially emit
an H$\alpha$ photon when they are produced in an excited state.
A fully description of the processes involved in the cosmic ray interactions
in the ISM to give account of the H$\alpha$ emission has to be carried out to confirm
our interpretation. In particular, the efficiency of each reaction
mentioned above and its probability to produce an H$\alpha$ photon is
unknown. The cosmic ray rate we have estimated from the H$\alpha$ surface
brightness is believed to be a conservative lower limit.

\section{Conclusions}

The analysis of the data shows the following:

\begin{enumerate}

\item  The H$\alpha$ emission is well correlated with the extinction map,
with an evident increase up to optical extinctions of $\sim$2 mag. 
There is a deep penetration of the ionisation flux that produces H$\alpha$ in regions 
with a molecular content, but also a likely influence of the ionisation 
flux on the balance of the chemical abundances in the densest regions. Cosmic 
rays of $\sim$ 100 MeV can reach the inner regions. 

\item The warm and cold components of large dust grains in LDN 1780 have 
been separated. The warm component is surrounding the cold component, and it is
mainly in the illuminated side of the cloud as the HI excess emission. The cold 
component is associated to the $^{13}$CO core.

\item The warm and cold components have an average colour temperature of 
25\,K and 16\,K (assuming $\beta$=2), respectively. The optical depth of the 
warm component is few tens lower than the value of the cold component. The observed 
colour temperatures are relatively high for moderate density regions (see Paper I), 
which support the presence of a large penetration depth ionisation source (i.e, 
cosmic rays). The cold component presents a ratio $\tau_{200}/A_V$ that varies
between 1\ 10$^{-4}$ and 3\ 10$^{-4}$ mag$^{-1}$, which are significantly
above the values of the warm component and around the value
of the diffuse interstellar medium (2.5 \ 10$^{-4}$ mag$^{-1}$).

\item The warm component consists of {\it warm} BGs that are 
influenced by the asymmetric radiation field, with has a major component from the 
Sco-Cen OB association and the Galactic plane. The UV radiation field, with a 
penetration depth much lower than the cosmic rays, might affect the 
outer regions of the warm component, which might be associated to the HI excess 
emission ($A_V$ up to 1 mag). The inner regions of the warm component could be 
partly associated with CO; the blue-shifted (out-flowing) CO component being 
likely affected by the UV radiation field, although $^{12}$CO correlates better 
with the cold component. The cold component consists of cold large dust grains 
that are shielded from the UV radiation field, but not from the incident cosmic 
ray flux.

\item The value of the observed ionisation rate $\xi$ $\sim$ 10$^{-16}$ 
$\gamma s^{-1}$ for LDN 1780 is relatively high in comparison with the values 
for dense regions (few$\times$ 10$^{-18}$ $\gamma s^{-1}$) and for the warm 
(T$\sim$8000\,K) and cold (T$\ge$50\,K) neutral medium ($\sim$1.8 10$^{-17}$ 
$\gamma s^{-1}$), but consistent with the enhancements found by McCall et al. 
(2002) in the diffuse ISM (few$\times$10$^{-15}$ $\gamma s^{-1}$).

\item Recently, Padoan \& Scalo (2005) have proposed the confinement of 
low-energy cosmic rays of $\sim$100 MeV by self generated MHD waves in the 
cloud, that can predict enhancements in the ionisation rate of up to 100 times the 
standard value for densities of $\sim$500 cm$^{-3}$. This value is consistent 
with the densities derived from CO observations in LDN 1780. The enhanced 
cosmic ray flux is likely associated to supernovae in the Sco-Cen OB association.

\end{enumerate}

\section*{Acknowledgements}

The authors thank Ren\'e Laureijs for providing the CO and $^{13}$CO data of 
LDN 1780. 

Based on observations with ISO, an ESA project  with instruments funded by ESA 
member states 
(especially the  PI countries France, Germany, the Netherlands and the United  
Kingdom) with 
participation of ISAS and NASA.

The ISOPHOT data presented in this paper were reduced using PIA, which is a 
joint development 
by the ESA Astrophysics Division and the ISOPHOT consortium, with the 
collaboration of the 
Infrared Analysis and Processing Center (IPAC) and the Instituto de Astrof\'\i 
sica de Canarias (IAC).

This publication makes use of data products from the Two Micron All Sky
  Survey, which is a joint project of the University of Massachusetts and the
  Infrared Processing and Analysis Center/California Institute of Technology,
  funded by the National Aeronautics and Space Administration and the
  National Science Foundation.

This research has made use of the NASA/IPAC Infrared Science Archive, which is 
operated by 
the Jet Propulsion Laboratory, California Institute of Technology, under 
contract with the 
National Aeronautics and Space Administration.

The Digitized Sky Survey was produced at the Space Telescope Science Institute 
under U.S. 
Government grant NAG W-2166. The images of these surveys are based on 
photographic data obtained 
using the Oschin Schmidt Telescope on Palomar Mountain and the UK Schmidt 
Telescope. The plates 
were processed into the present compressed digital form with the permission of 
these institutions.

\end{document}